\documentclass[iop]{emulateapj}

\def\xrt{{{\it Swift}-XRT}\/}
\def\fer{{{\it Fermi}}\/}

\shorttitle{Identifying the 3FHL catalog: Results of the KOSMOS optical spectroscopy campaign}
\shortauthors{Marchesi et al.}

\usepackage{natbib}
\usepackage{float}
\usepackage{color}
\usepackage{graphicx}
\usepackage{gensymb}
\usepackage{array}
\usepackage{enumitem}
\usepackage{hyperref}

\begin{document}

\title{Identifying the 3FHL catalog: II. Results of the KOSMOS optical spectroscopy campaign}

\author{S. Marchesi\altaffilmark{1}, A. Kaur\altaffilmark{1}, M. Ajello\altaffilmark{1}} 

\altaffiltext{1}{Department of Physics and Astronomy, Clemson University, Clemson, SC 29634, USA}

\begin{abstract}
We present the results of the optical spectroscopy follow-up of a sample of 28 unclassified blazars from the Third \fer--LAT Catalog of High-Energy Sources (3FHL). All the spectra were taken with the 4m Mayall telescope at Kitt Peak. With this follow-up program we are able to classify 27 out of 28 objects as BL Lacs, while the remaining one is a flat spectrum radio quasar. We determine a redshift ($z$) for three of these objects and a lower limit on $z$ for other four sources: the farthest object for which we obtain a redshift has $z$$>$0.836. These results are part of a more extended campaign of optical spectroscopy follow-up of 3FHL blazars, aimed to obtain a complete sample of blazars at $>$10\,GeV which will then be used to extend our knowledge on blazar emission mechanisms and on the extragalactic background light.
\end{abstract}


\section{Introduction}
The Large Area Telescope \citep[LAT;][]{atwood09}, mounted on the {\it Fermi Gamma-ray Space Telescope}, represented a breakthrough for the study and understanding of the high-energy Universe, and in particular of blazars, the most extreme class of active galactic nuclei and the most numerous population detected in $\gamma$-rays. The Third \fer--LAT Catalog of High-Energy Sources \citep[3FHL][]{ajello17}, based on seven years of \fer\ observations, is the latest catalog of sources detected at $>$10\,GeV and fully exploits the improvement in performances delivered by `Pass~8'', the newest event-level analysis method \citep{atwood13}. The 3FHL reports the detection of more than 1500 sources all-sky, a dramatic improvement of a factor of 3 over its predecessor, the 1FHL\footnote{The 2FHL catalog is a catalog of sources detected above 50\,GeV.} catalog \citep{ackermann16a}. The 3FHL represents the current deepest look at the very high-energy sky and in all likelihood it will be used to plan most of the observations of the upcoming Cherenkov Telescope Array  \citep[CTA;][]{cta17}.

To maximize the scientific impact of the 3FHL catalog, the sample should be completely identified (i.e., the nature of all sources should be known) and the vast majority of the objects should have a redshift ($z$). At the present day, however, while the 3FHL catalog contains more than 1500 sources detected at $>$10\,GeV, 1227 of which are of confirmed extragalactic origin, only a minority of them (42.5\,\%) have a measured $z$ and a significant fraction ($\sim$25\%) lacks both a redshift and a secure classification. Particularly, the 3FHL contains 290 blazar candidates of uncertain type (BCUs) and 177 unassociated sources. Spectroscopic observations of most of these sources are needed to fully realize the scientific potential of the 3FHL catalog. The redshift measurement is particularly important, since it allows us to use the 3FHL catalog to investigate a variety of topics, such as the blazars energetics and emission mechanism \citep[see, e.g.,][]{ghisellini17}; the extragalactic background light (EBL), i.e., the integrated emission of all stars and galaxies in the Universe, and its evolution with redshift \citep{ackermann12,abramowski12,dominguez13}; the role of blazars in accelerating cosmic rays \citep[see, e.g.,][]{furniss13}; to study the evolution of blazars across cosmic time \citep{ajello14}.

At the present day, spectroscopic follow up of unclassified blazars have been mainly performed with 4m- or 8m-class telescopes \citep[e.g.,][]{sbaruffati05,sbaruffati06a,shaw13,massaro14,paggi14,landoni15,ricci15,alvarez16a,alvarez16b,marchesini16,pena17}. These works showed that 4m telescopes are effective in distinguish between flat spectrum radio quasars (FSRQs) and BL Lacs. FSRQs present strong emission features and their redshift are therefore easily measured; BL Lacs, instead, are typically characterized by narrow \citep[equivalent width, EW$\leq$5--20\AA; see, e.g., ][]{marcha96} emission lines, although this EW-based classification is phenomenological, rather than physical, and several cases of  ``transition blazars'', i.e., objects with significant EW variability, mostly BL Lacs showing broad permitted emission lines, are reported in the literature \citep[see, e.g.,][]{ulrich81,corbett00,ghisellini11,shaw12,ruan14}.
Furthermore, BL Lacs can be divided in three different classes, based on the frequency of the synchrotron peak: sources with log$_{10}$($\nu_{\rm peak}^{\rm S}$/Hz)$<$14 are classified as low-synchrotron peak (LSP), sources with 14$<$log$_{10}$($\nu_{\rm peak}^{\rm S}$/Hz)$<$15 are classified as intermediate-synchrotron peak (ISP) and sources with log$_{10}$($\nu_{\rm peak}^{\rm S}$/Hz)$>$15 are classified as high-synchrotron peak (HSP). Historically, it has been possible to obtain a spectroscopic redshift for 67\%, 37\% and 48\% of LSP (including FSRQs), ISP and HSP objects, respectively \citep{shaw12,shaw13}.

In this paper, we report the results of a spectroscopic follow-up campaign with the 4 meter Mayall telescope at the Kitt Peak National Observatory (KPNO) of 28 3FHL sources selected among those lacking of redshift information. The work is organized as follows: in Section \ref{sec:sample} we present the sample selection criteria. In Section \ref{sec:analysis} we describe the observing process and the data analysis. In Section \ref{sec:results} we report the results of the spectroscopic campaign, both in general terms and in detail for each of the 28 sources. Finally, we report our conclusions in Section \ref{sec:concl}.

\section{Sample selection}\label{sec:sample}
Our long term goal is to provide redshifts and type information for all the 706 sources with no redshift or classification in the 3FHL catalog. In this work we report the results of the first step of this ambitious campaign, i.e., the observation of a subsample of 28 sources with the Mayall 4m telescope at KPNO\footnote{These observations were awarded as part of the Fermi-GI cycle 10, accepted proposal 101287, PI: S. Marchesi.}. Out of this 28 sources, 23 are blazar candidate of uncertain type \citep[BCUs;][]{ackermann15}. BCUs have multiwavelength properties which allow to reliably classify them as blazars, however their low-quality optical spectra prevent us to distinguish their class, i.e., either BL Lacs or flat spectrum radio quasars (FSRQs). 

\begin{figure*}
\begin{minipage}[b]{.5\textwidth}
  \centering
  \includegraphics[width=0.8\textwidth]{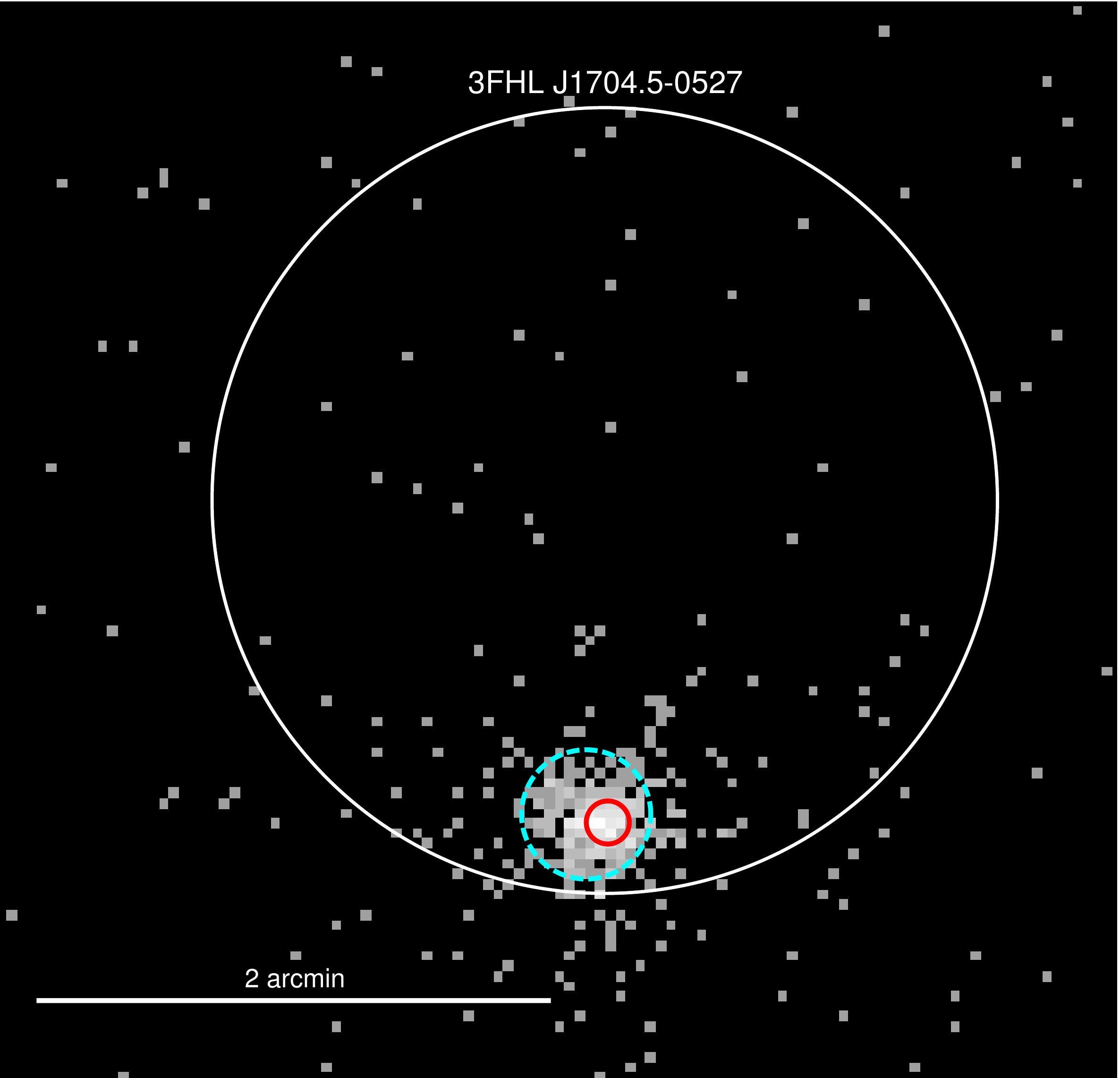}
  \end{minipage}
\begin{minipage}[b]{.5\textwidth}
  \centering
  \includegraphics[width=0.8\textwidth]{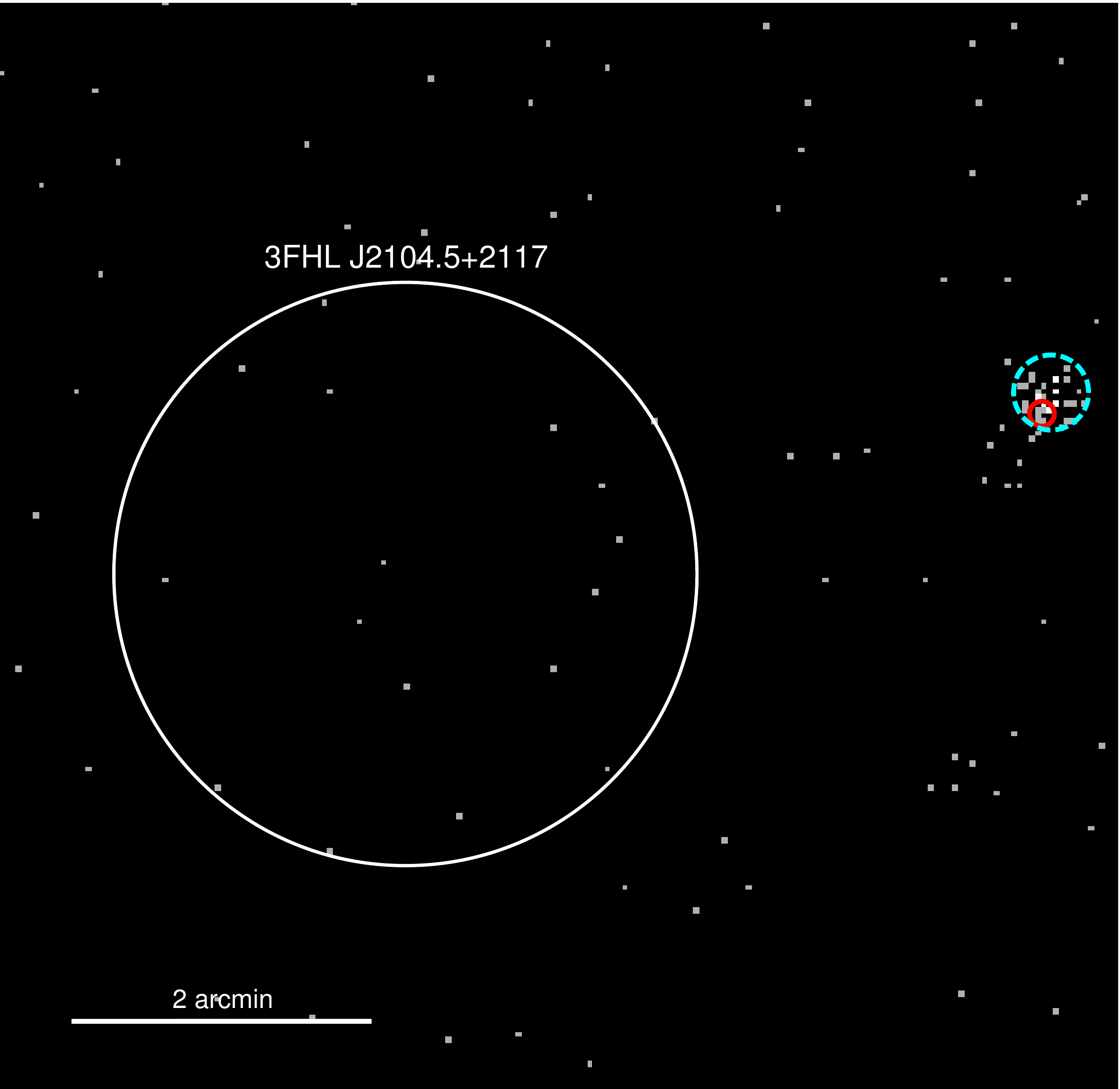}
  \end{minipage}
\begin{minipage}[b]{.5\textwidth}
  \centering
  \includegraphics[width=0.8\textwidth]{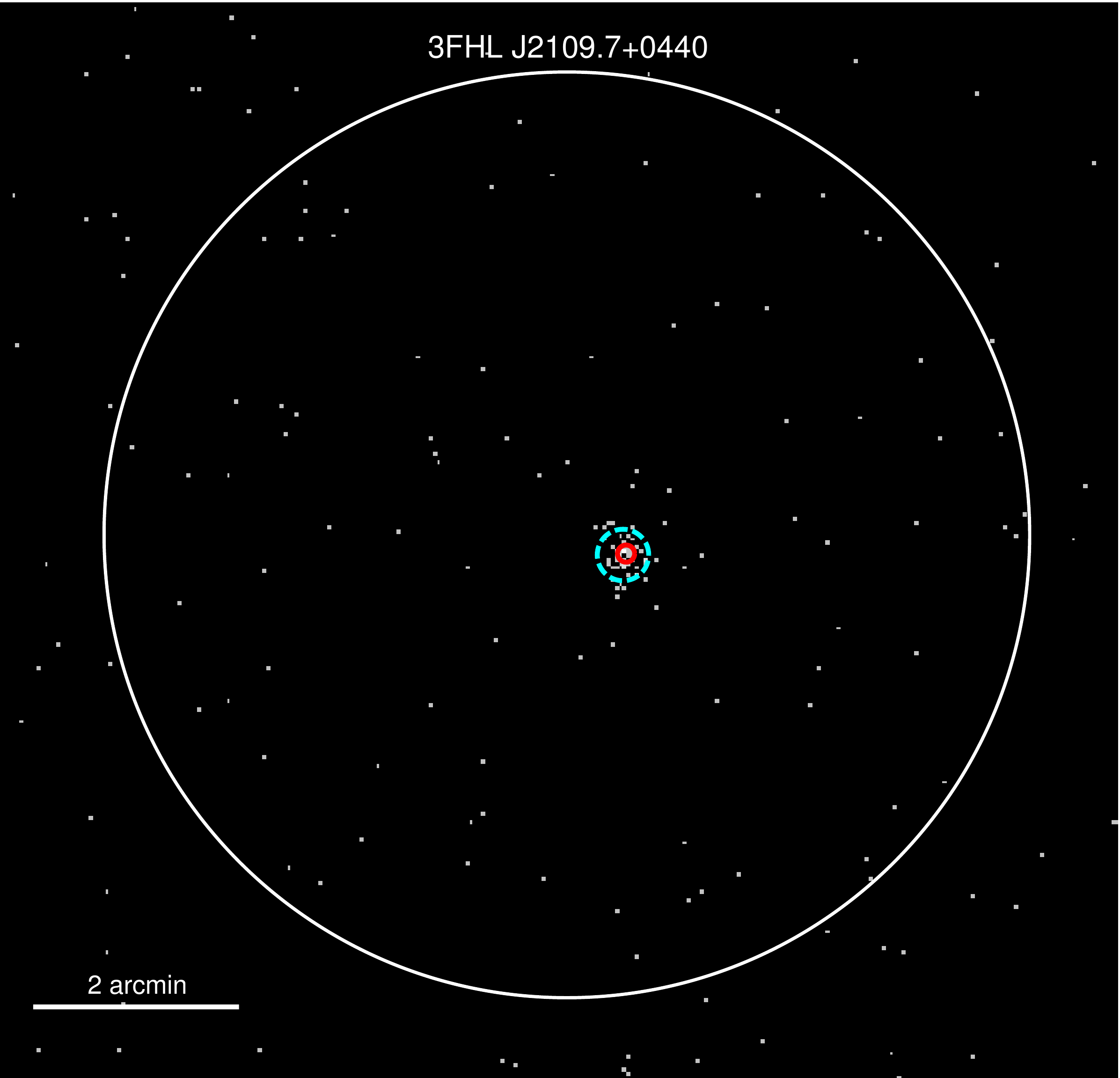}
  \end{minipage}
\begin{minipage}[b]{.5\textwidth}
  \centering
  \includegraphics[width=0.8\textwidth]{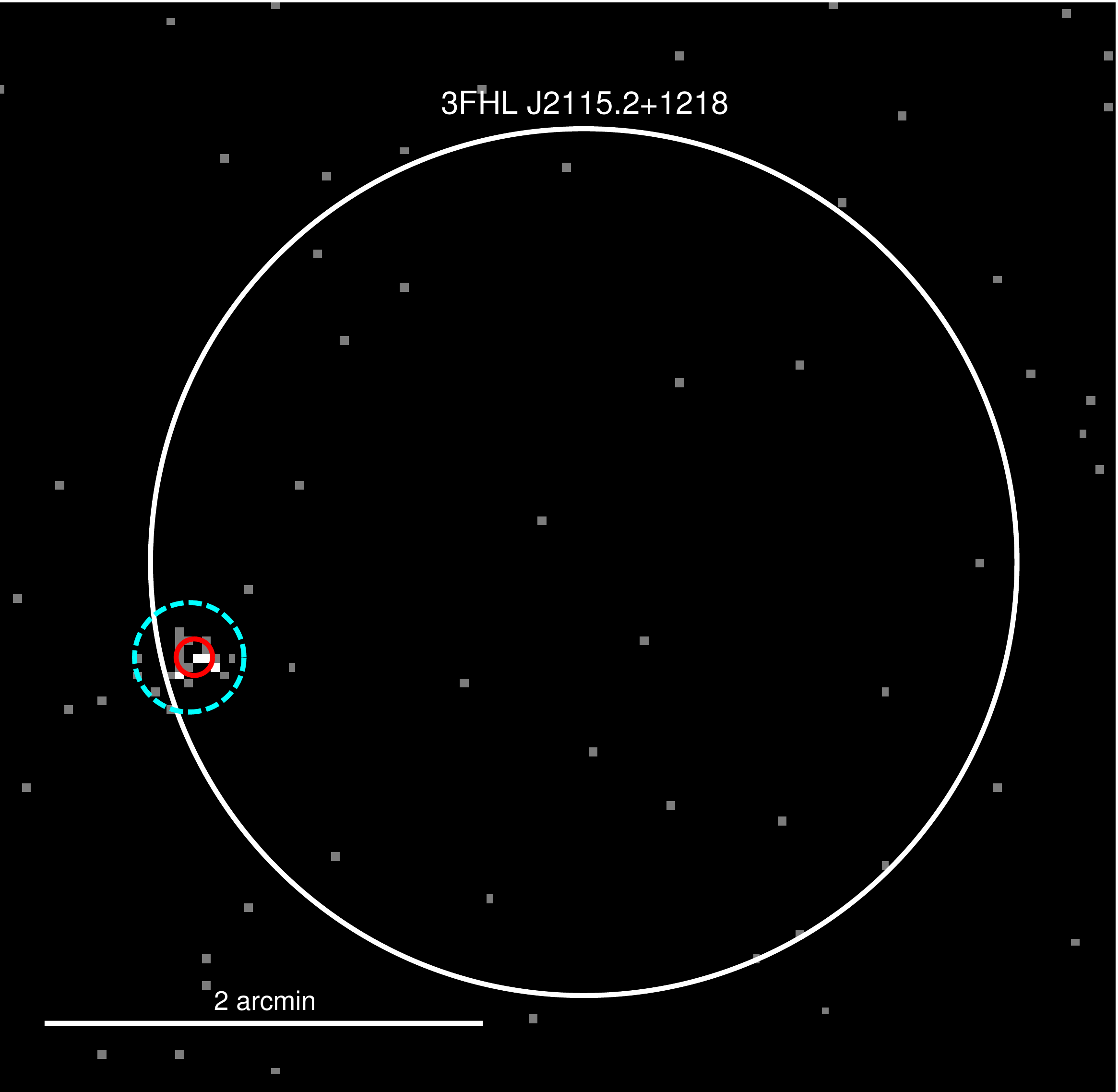}
  \end{minipage}
  \begin{minipage}[b]{.5\textwidth}
  \centering
  \includegraphics[width=0.8\textwidth]{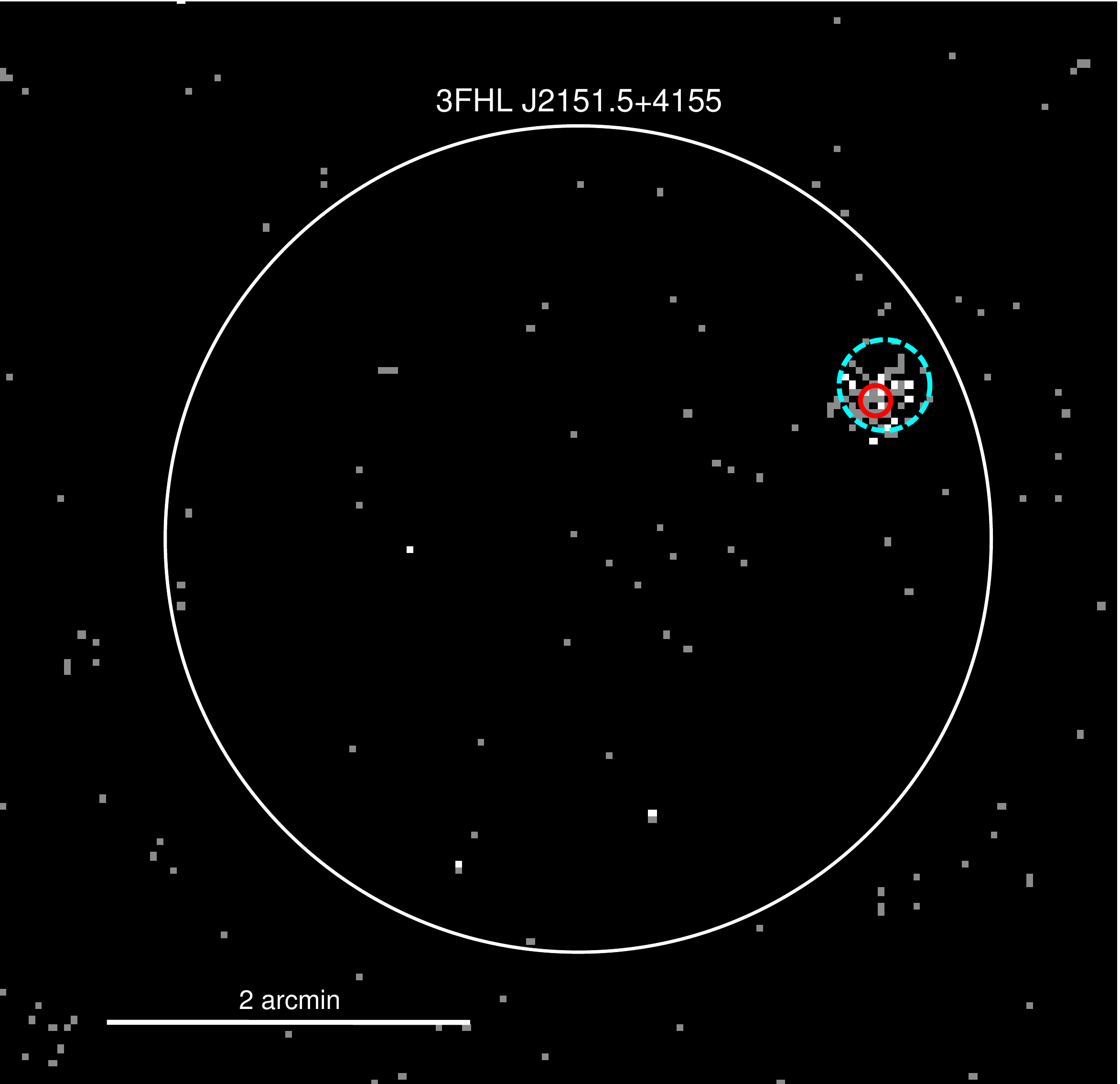}
  \end{minipage}
\caption{\xrt\ 0.5--10\,keV images of the five ``unassociated'' objects in our sample. The LAT 95\% confidence position uncertainty is shown as a white solid circle. The \xrt\ and NVSS positions are plotted as a 15$^{\prime\prime}$ cyan dashed circle and as a 5$^{\prime\prime}$ red solid circle, respectively. As can be seen, all 3FHL sources have a single, bright \xrt\ counterpart within the \xrt\ field of view.}\label{fig:xrt_unassociated}
\end{figure*}

Other five objects are instead classified as ``unassociated'', i.e., in 3FHL catalog they are not associated with a counterpart. These objects are part of a larger sample which we will present in a companion paper (Kaur et al. submitted). In this work, we look for potential 0.5--10\,keV counterparts of the 3FHL sources using archival \xrt\ observations: 
the XRT position accuracy of a few arcseconds allows one to easily identify the optical and/or radio counterparts for these objects. In Figure \ref{fig:xrt_unassociated} we report the \xrt\ 0.3--10\,keV images of the five unassociated objects in our sample: as can be seen, four out of five 3FHL sources have a single, bright (count rate in the 0.5--10\,keV band $r$$\gtrsim$0.01 cts s$^{-1}$)) counterpart within the LAT 95\% confidence positional uncertainty. The only exception is 3FHL J2104.5+2117, where no significant 0.3--10\,keV emission is observed within the LAT source region. However, a X-ray source associated with a radio counterpart is located at $\sim$4.5$^{\prime}$ from the LAT centroid, and we thus choose this object (radio source NVSS J210415+211805) as the most likely counterpart.

Besides their lacking of redshift and type information, we selected the sources in our sample with the following criteria:
\begin{enumerate}
\item Having available optical magnitude, and the magnitude being $V$$\leq$21.5.
\item Being bright in the hard $\gamma$-rays ($f_{\rm 50-150 GeV}>$10$^{-13}$ erg s$^{-1}$ cm$^{-2}$). Selecting 3FHL objects bright in the 50--150\,GeV band ensures that the completeness of the 3FHL catalog evolves to lower fluxes as more optical observations are performed.
\item Being observable from Kitt Peak (Dec$>$--20\,\degree) and during our observations (in August, this implies R.A.$\geq$17h00m00s and R.A.$\leq$3h00m00s; in October, it is R.A.$\geq$18h00m00s and R.A.$\leq$7h00m00s).
\end{enumerate}

66 3FHL sources satisfy all these criteria: 61 BCUs and five ``unassociated'' sources studied in Kaur et al. (submitted). Our 28 sources were selected among these 66 objects with the goal of covering a wide range of optical magnitudes (V=[14--20]) and, consequently, of potential redshifts and luminosities. We report in Table \ref{tab:sample} a summary of the sources analyzed in this work. 

\begingroup
\renewcommand*{\arraystretch}{1.8}
\begin{table*}
\centering
\scalebox{0.95}{
\begin{tabular}{ccccccccc}
\hline
\hline
3FHL Name & Counterpart & R.A. & Dec & Type & mag & E(B-V) & Obs Date & Exposure\\
(1) & (2) & (3) & (4) & (5) & (6) & (7) & (8) & (9)\\
\hline
J0031.2+0727 & NVSS J003119+072456 & 00:31:19.7 & +07:24:53.6 &   BCU &   19.2 & 0.03 & 2017 Oct 09 & 4800 \\
J0040.3+4049 & NVSS J004013+405005 & 00:40:13.8 & +40:50:04.7 &  BCU &    19.2 & 0.20 & 2017 Oct 11 & 6000 \\
J0127.2+0325 & NVSS J012713+032259 & 01:27:13.9 & +03:22:59.0 &   BCU &   19.5 & 0.02 & 2017 Oct 10 & 4500  \\
J0134.4+2638 & NVSS J013427+263842 & 01:34:28.2 & +26:38:43.0 &   BCU &   16.4 & 0.09 & 2017 Oct 10 & 1700 \\
J0241.3+6543 & NVSS J024121+654311  & 02:41:21.7 & +65:43:12.0 &   BCU &   19.7  & 1.09 & 2017 Oct 12 & 6800 \\
J0305.2--1609 & PKS 0302--16                   & 03:05:15.3 & --16:08:12.1 &  BCU &   18.2 & 0.04 & 2017 Oct 12 & 3600 \\
J0353.4+8256 & NVSS J035309+825631 & 03:53:09.5 & +82:56:31.3 &   BCU &   14.3 & 0.10 & 2017 Oct 11 & 900 \\
J0420.2+4011 & NVSS J042013+401122 & 04:20:13.4 & +40:11:21.7 &   BCU &   17.6 & 0.60 & 2017 Oct 10 &  3900 \\
          ...           &               ...                      &      ...          &        ...          &      ...   &      ...  &  ...  & 2017 Oct 11 &  1800 \\
J0431.8+7403 & NVSS J043145+740327 & 04:31:45.2 & +74:03:27.6 &   BCU &   18.6  & 0.15 & 2017 Oct 12 & 3900 \\
J0434.7+0921 & NVSS J043440+092348 & 04:34:40.9 & +09:23:48.5 &   BCU &   18.1 & 0.22 & 2017 Oct 12 & 3200  \\
J0640.0--1254 & NVSS J064007--125315  & 06:40:07.2 & --12:53:15.6 &  BCU &   15.2 & 0.38 & 2017 Oct 11 & 1140 \\
J1704.5--0527 & NVSS J170433--052839* & 17:04:33.8 & --05:28:39.8 & --  &   18.0 & 0.46 & 2017 Aug 11 & 3600 \\
J1820.4+3623 & NVSS J182021+362343 & 18:20:21.0 & +36:23:43.3 &  BCU &   18.8 & 0.03 & 2017 Oct 11 & 4800 \\
J1841.3+2909 & NVSS J184121+290945 & 18:41:21.6 & +29:09:45.3 &   BCU &   17.1  & 0.21 & 2017 Oct 09 & 3600 \\
J1850.4+2631 & NVSS J185023+263151 & 18:50:24.0 & +26:31:51.3 &   BCU &   18.1 & 0.18 & 2017 Oct 11 & 3800 \\
J1904.1+3627 & NVSS J190411+362700 & 19:04:11.9 & +36:26:58.9 &   BCU &   13.7 & 0.08 & 2017 Oct 09 & 1500 \\
J1911.5--1908 & NVSS J191129--190823             & 19:11:29.7 & --19:08:22.0 &  BCU &   15.3 & 0.14 & 2017 Oct 12 &  2160 \\
J1949.5+0906 & 2MASSJ19493419+0906537 & 19:49:34.2 & +09:06:53.7 &  BCU  &  17.4 & 0.18 & 2017 Oct 12 & 3200 \\
J2024.4--0847  & NVSS J202429--084804 &  20:24:29.4 & --08:48:04.8 & BCU &   17.9  & 0.07 & 2017 Oct 12 & 3000 \\
J2104.5+2117 & NVSS J210415+211805* & 21:04:15.9 & +21:18:05.2 & --   &  20.0   & 0.15 & 2017 Aug 13 & 6000 \\
J2109.7+0440 & NVSS J210939+044000* & 21:09:40.1 & +04:40:00.4 & --   &  16.8   & 0.10 & 2017 Aug 12  & 2400 \\
J2115.2+1218 & NVSS J211522+121802* & 21:15:22.0 & +12:18:02.9 & --   &  16.3   & 0.06 & 2017 Aug 13 & 4800 \\
J2151.5+4155 & NVSS J215122+415632* & 21:51:23.0 & +41:56:32.9 & --   &  13.6   & 0.28 & 2017 Aug 13 & 2100 \\
J2212.6+2759 & NVSS J221239+275937 & 22:12:39.1 & +27:59:38.5 &  BCU &   18.1 & 0.07 & 2017 Oct 09 & 4200 \\
J2220.5+2813 & NVSS J222028+281355 & 22:20:28.7 & +28:13:55.6 &  BCU &   15.4 & 0.08 & 2017 Oct 10 & 900 \\
          ...           &               ...                      &      ...          &        ...          &     ...   &      ...  &  ...  & 2017 Oct 11 &  2100 \\
J2245.9+1545 & NVSS J224604+154437 & 22:46:05.0 & +15:44:35.4 &  BCU &   19.0 & 0.07 & 2017 Oct 10 & 4000 \\
J2300.0+4054 & NVSS J230012+405224 & 23:00:12.5 & +40:52:24.3 &  BCU &   17.4 & 0.10 & 2017 Oct 11 & 4800 \\
J2358.5+3829 & NVSS J235825+382857 & 23:58:25.2 & +38:28:57.0 &  BCU &   17.5 & 0.11 & 2017 Oct 11 & 3600 \\
\hline

\hline
\hline
\end{tabular}}\caption{Properties and summary of the observations of the 28 sources analyzed in this work. (1): source name in the 3FHL catalog \citep{ajello17}. (2): optical counterpart. Sources flagged with * are objects for which the optical counterpart has been identified through a \xrt\ follow-up (Kaur et al. 2018 in prep.). (3) Right ascension. (4) Declination. (5) Source classification: ``BCU'' are blazars of uncertain type, the sources with no classification are those for which we found a \xrt\ counterpart. (6) V band magnitude. (7) $E(B-V)$, as obtained from the NASA/IPAC Infrared Science Archive online tool, using the measurements of \citet{schlafly11}. (8) Date of observation. (9) Exposure time (in seconds). }
\label{tab:sample}
\end{table*}
\endgroup

\section{Observations and data analysis}\label{sec:analysis}
All the spectra analyzed in this work were acquired with the KOSMOS spectrograph mounted on the KPNO Mayall 4m \citep{martini14}, using the Blue VPH grism in conjunction with the 3500--6200\,\AA, 0.9$^{\prime\prime}$ slit. This experimental setup corresponds to a dispersion of $\sim$ 4\,\AA\ pixel$^{-1}$ and a spectral resolution R$\sim$2100. The data were taken with the slit aligned along the parallactic angle. The observations were performed in seven ``grey'' nights of observations, three nights in August, 2017 and four in October, 2017: however, due to poor weather conditions and technical issues, the effective number of nights of observations is $\sim$4.5. The observing conditions were spectroscopic. All the observations were performed remotely. 

All sources have been observed at least three times\footnote{3FHL J2109.7+0440 have been observed three times, but one of the observations was interrupted due to bad weather conditions, and we do not use it in our analysis} and the single observations have then been combined to reduce cosmic rays contribution and instrumental effects. The data reduction and spectral extraction procedure has been carried out using a standard IRAF pipeline \citep{tody86}, with bias subtraction, flat field normalization and bad pixel correction. All spectra have also been visually inspected to remove any artificial feature.

The spectra have been wavelength calibrated using Iron-Argon lamps, whose spectra were acquired either before or after each source observation, to take into account possible shifts in the pixel-$\lambda$ calibration, due to changes in the telescope position during the night. The spectra have then been flux calibrated using a spectrophotometric standard: in each observing night we observed two spectrophotometric standards, one at the beginning and the other at the end of the night: the standards were observed using the same 0.9$^{\prime\prime}$ slit used in the rest of the analysis. Finally, each spectrum has been corrected for Galactic reddening, using the \citet{cardelli89} extinction law and the $E(B-V)$ value obtained from the NASA/IPAC Infrared Science Archive online tool\footnote{\url{https://irsa.ipac.caltech.edu/applications/DUST/}}: we adopted the measurements of \citet{schlafly11}.

\section{Spectral analysis results}\label{sec:results}
We report in Table \ref{tab:results} a summary of the spectral analysis results. The spectra signal-to-noise ratios (S/N) have been computed on the normalized spectra. 
The S/N has then been measured in five regions, each of which having size $\Delta$$\lambda$$\sim$50\,\AA, chosen in featureless parts of the spectrum in the energy range 4500--6000\,\AA.

\begin{figure*}
\begin{minipage}[b]{.5\textwidth}
  \centering
  \includegraphics[width=0.9\textwidth]{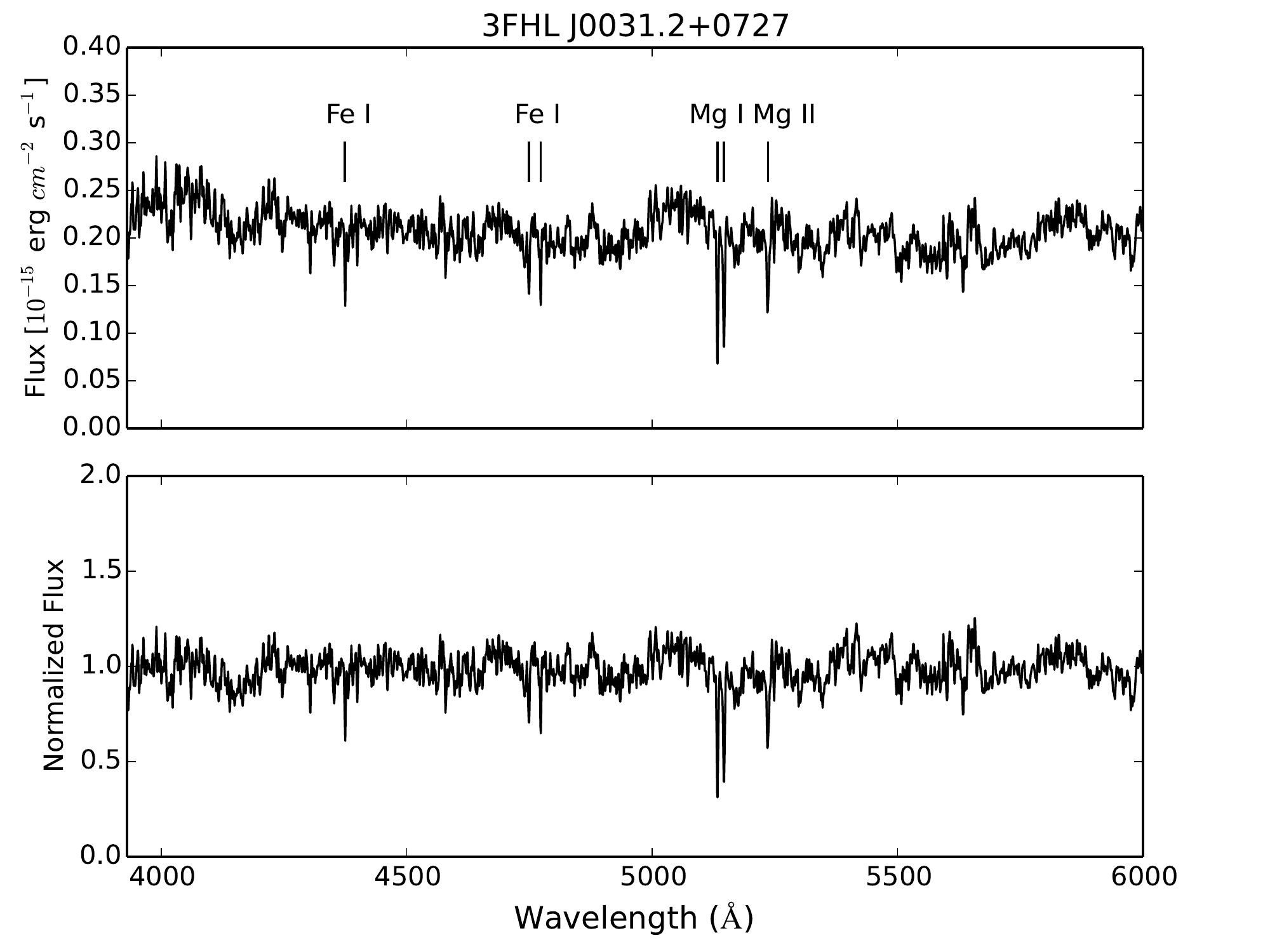}
  \end{minipage}
\begin{minipage}[b]{.5\textwidth}
  \centering
  \includegraphics[width=0.9\textwidth]{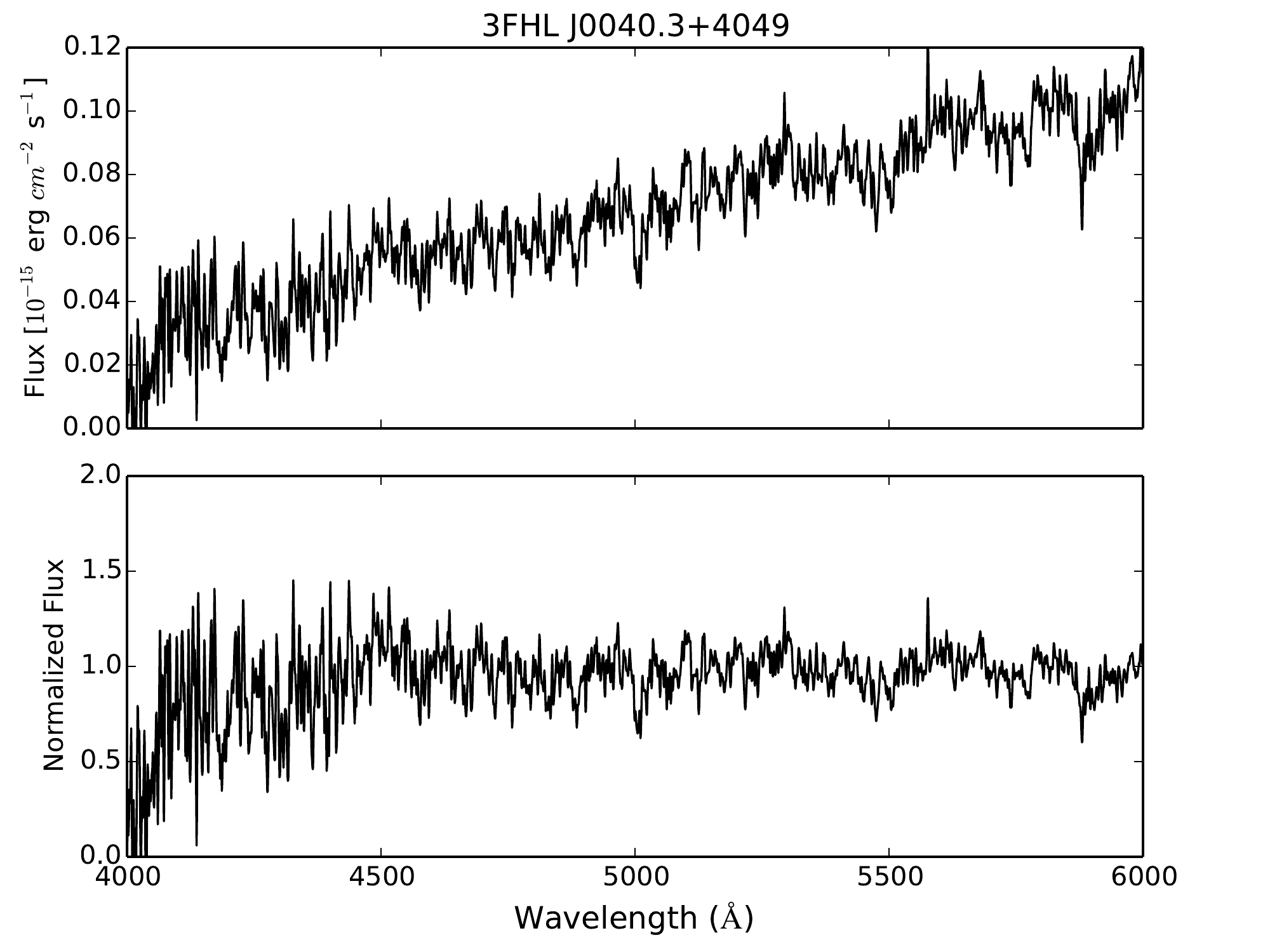}
  \end{minipage}
\begin{minipage}[b]{.5\textwidth}
  \centering
  \includegraphics[width=0.9\textwidth]{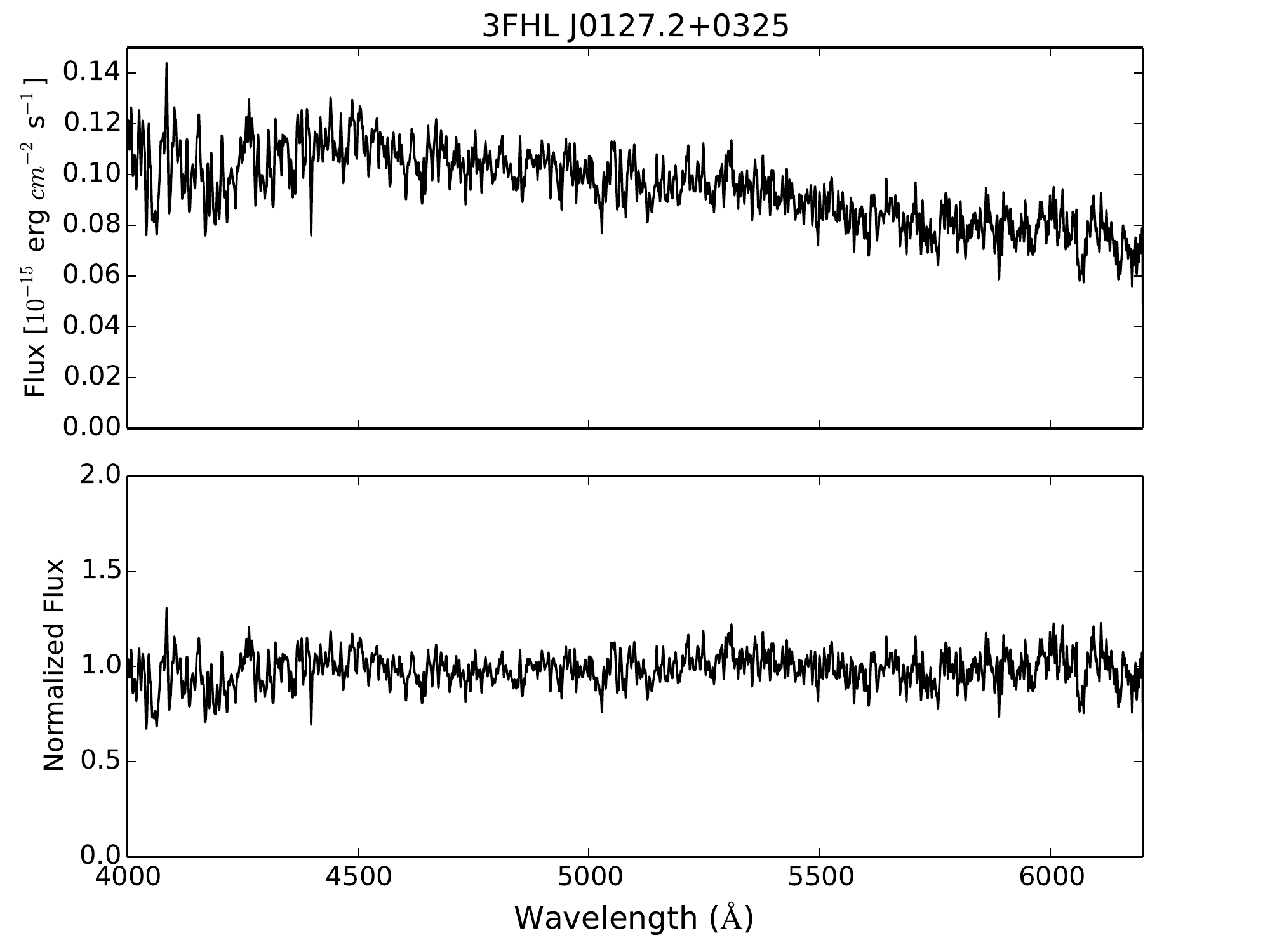}
  \end{minipage}
\begin{minipage}[b]{.5\textwidth}
  \centering
  \includegraphics[width=0.9\textwidth]{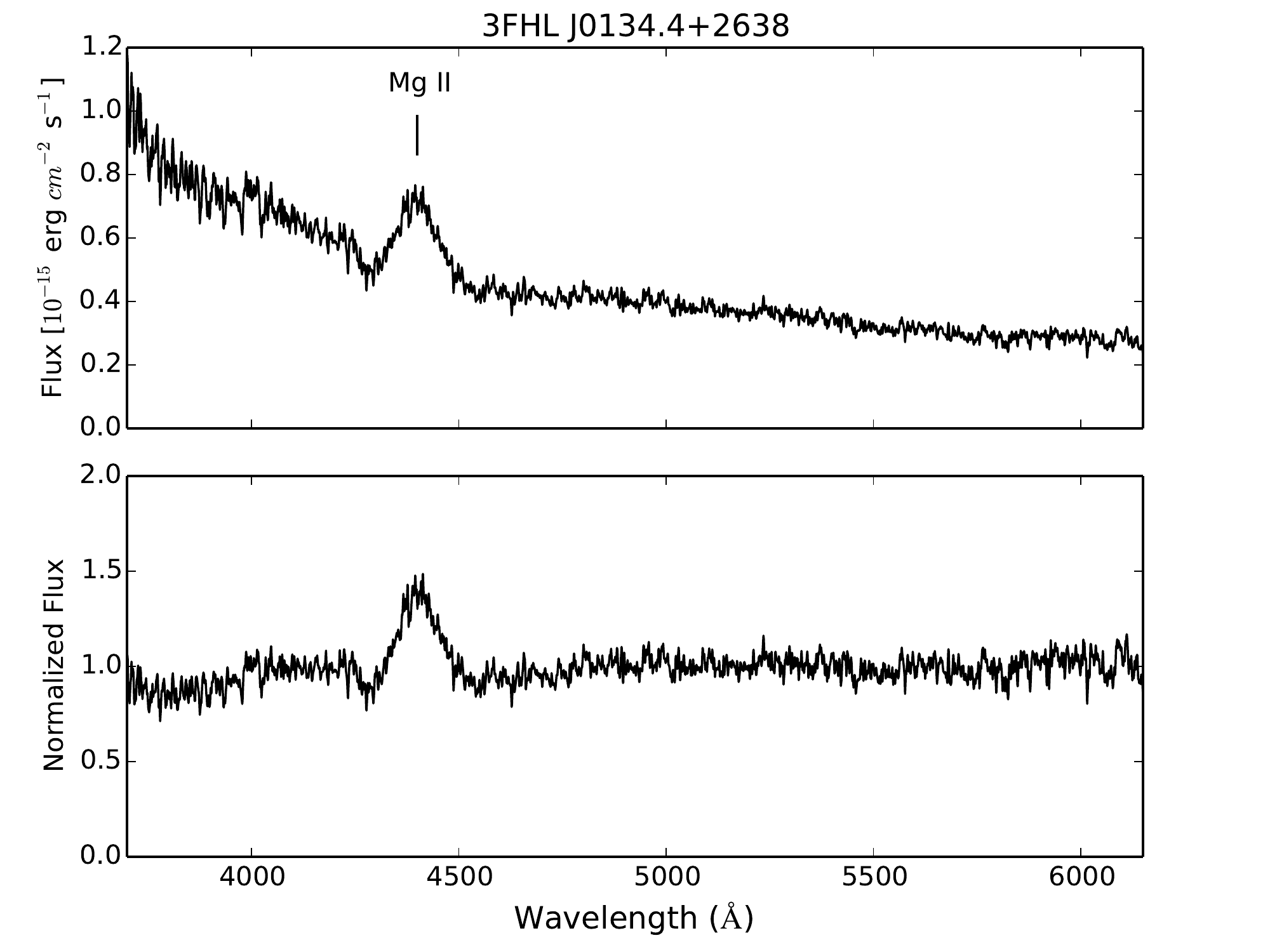}
  \end{minipage}
    \vspace{0.5cm}
\begin{minipage}[b]{.5\textwidth}
  \centering
  \includegraphics[width=0.9\textwidth]{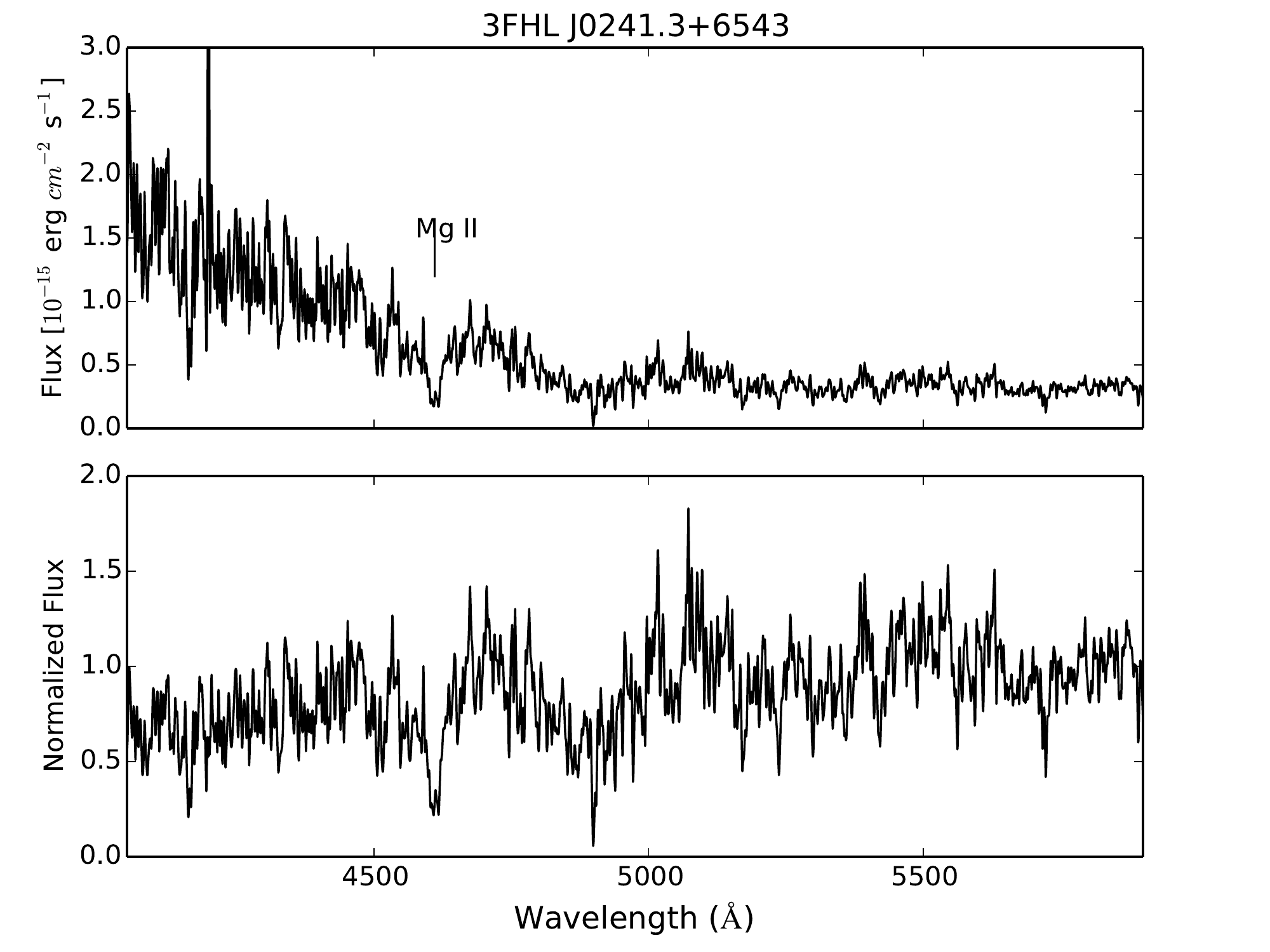}
  \end{minipage}
  \begin{minipage}[b]{.5\textwidth}
  \centering
  \includegraphics[width=0.9\textwidth]{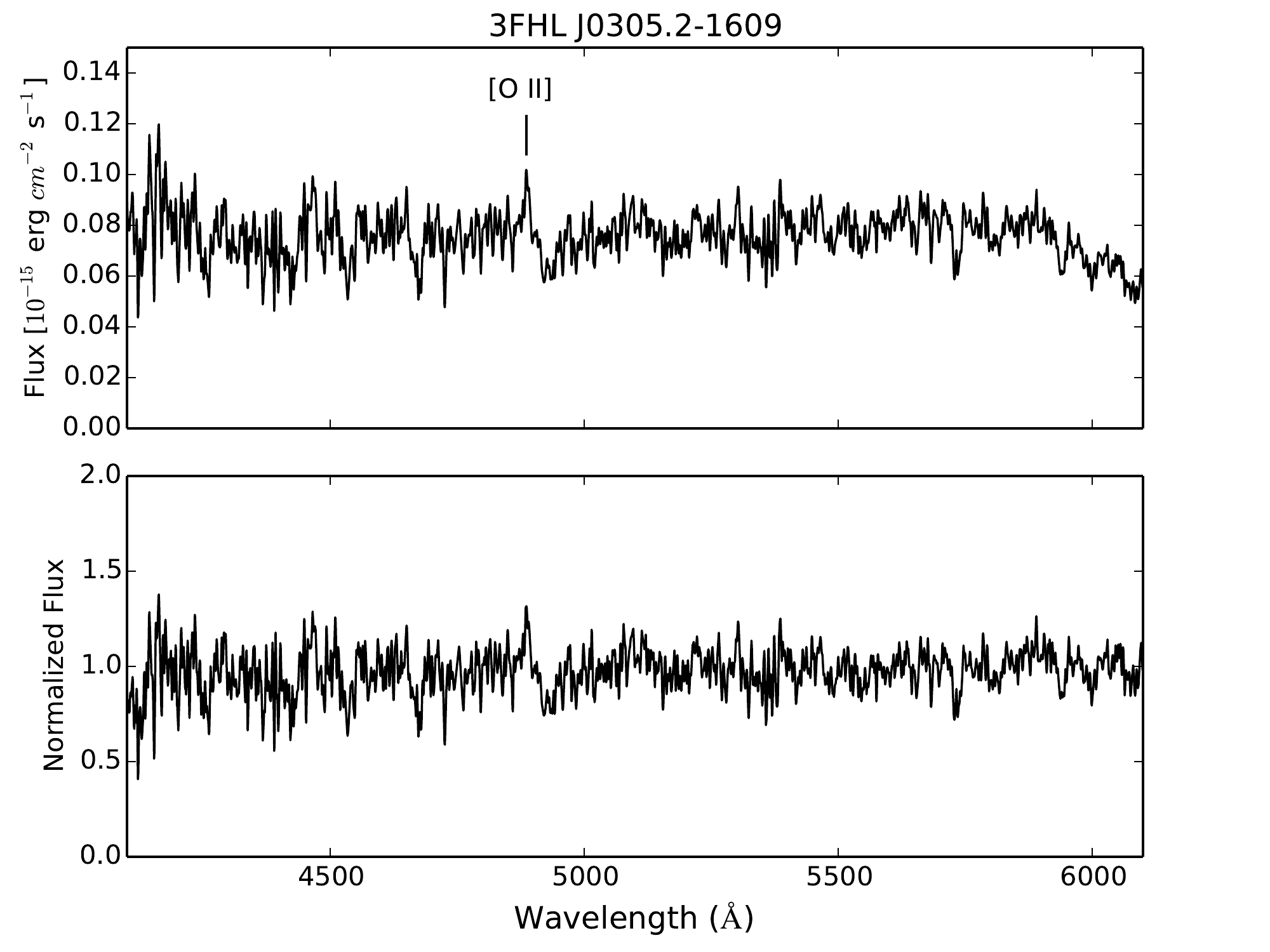}
  \end{minipage}
\caption{Optical spectra of six out of 28 objects in our sample. The spectra have been smoothed for visualization purposes.}\label{fig:spectra1}
\end{figure*}

\begin{figure*}
\begin{minipage}[b]{.5\textwidth}
  \centering
  \includegraphics[width=0.9\textwidth]{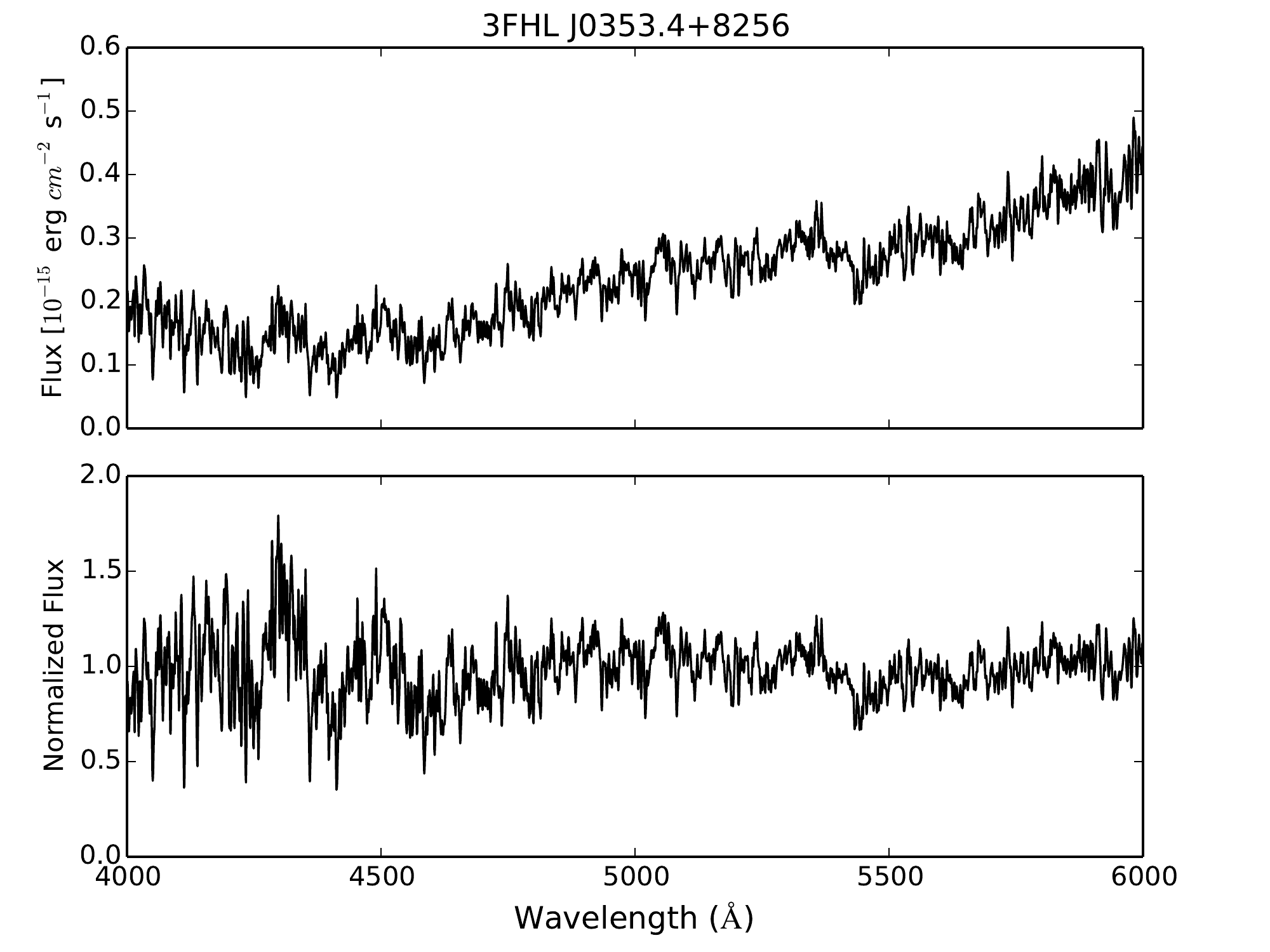}
  \end{minipage}
\begin{minipage}[b]{.5\textwidth}
  \centering
  \includegraphics[width=0.9\textwidth]{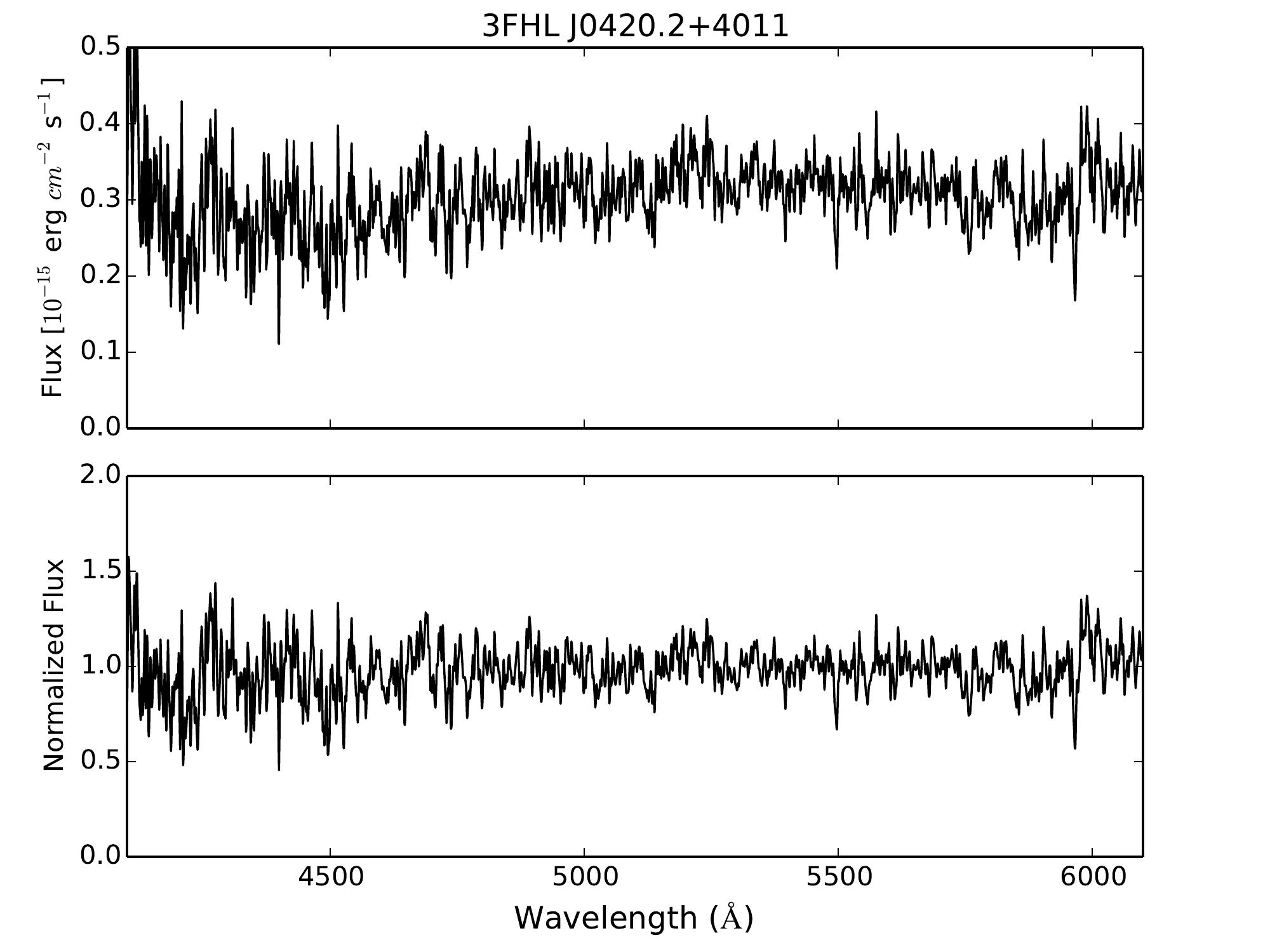}
  \end{minipage}
    \vspace{0.5cm}
\begin{minipage}[b]{.5\textwidth}
  \centering
  \includegraphics[width=0.9\textwidth]{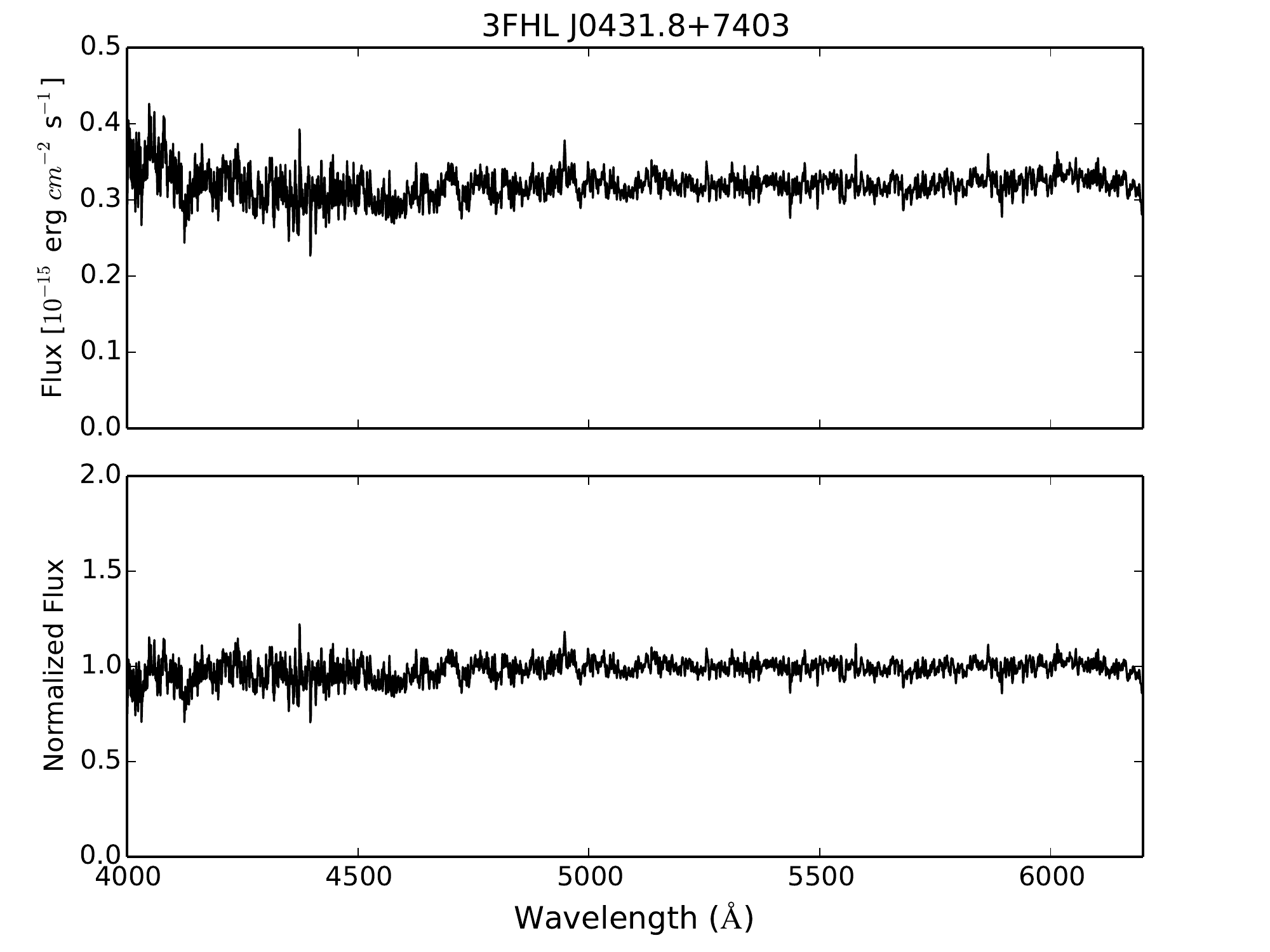}
  \end{minipage}
\begin{minipage}[b]{.5\textwidth}
  \centering
  \includegraphics[width=0.9\textwidth]{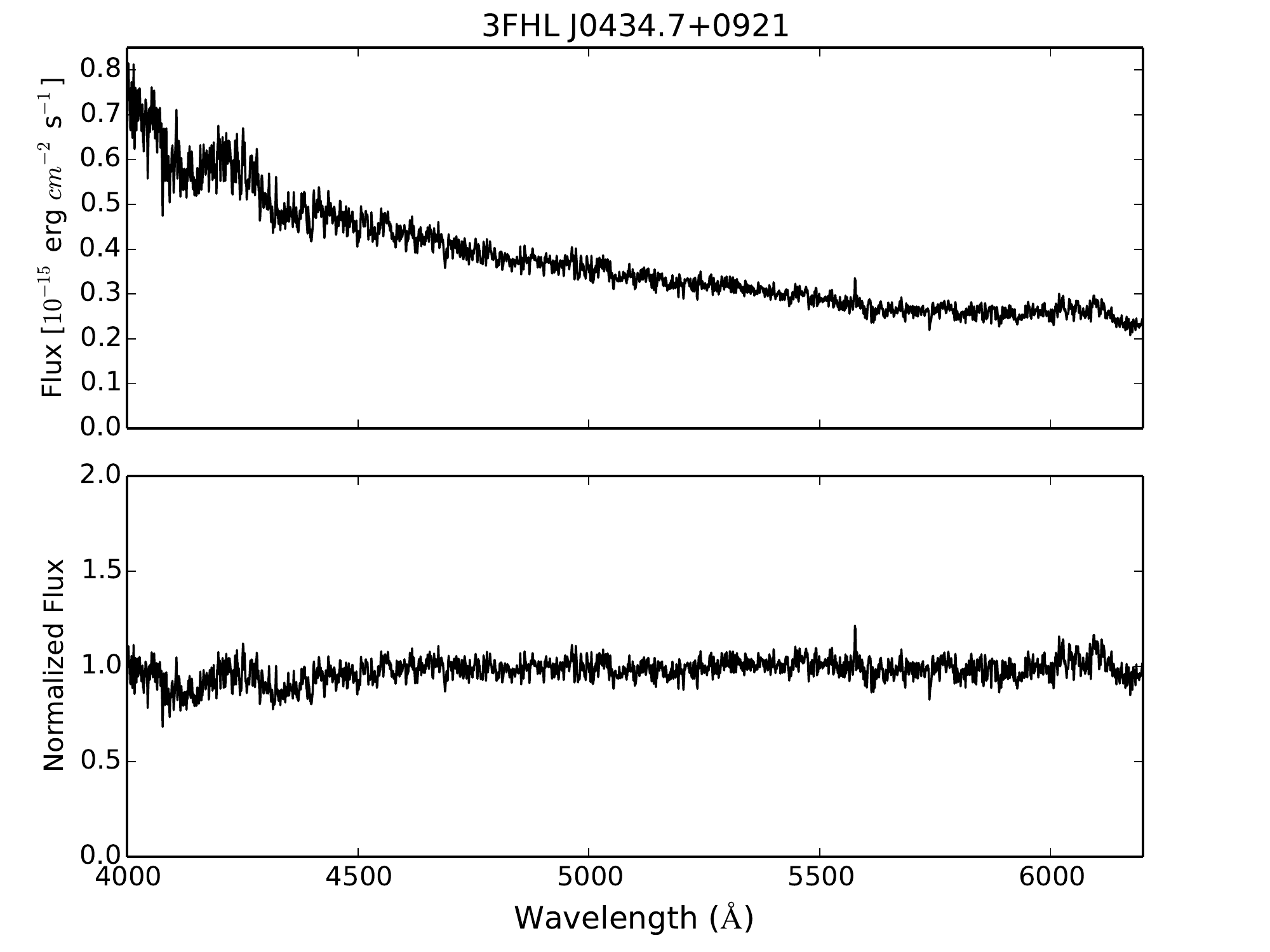}
  \end{minipage}
    \vspace{0.2cm}
\begin{minipage}[b]{.5\textwidth}
  \centering
  \includegraphics[width=0.9\textwidth]{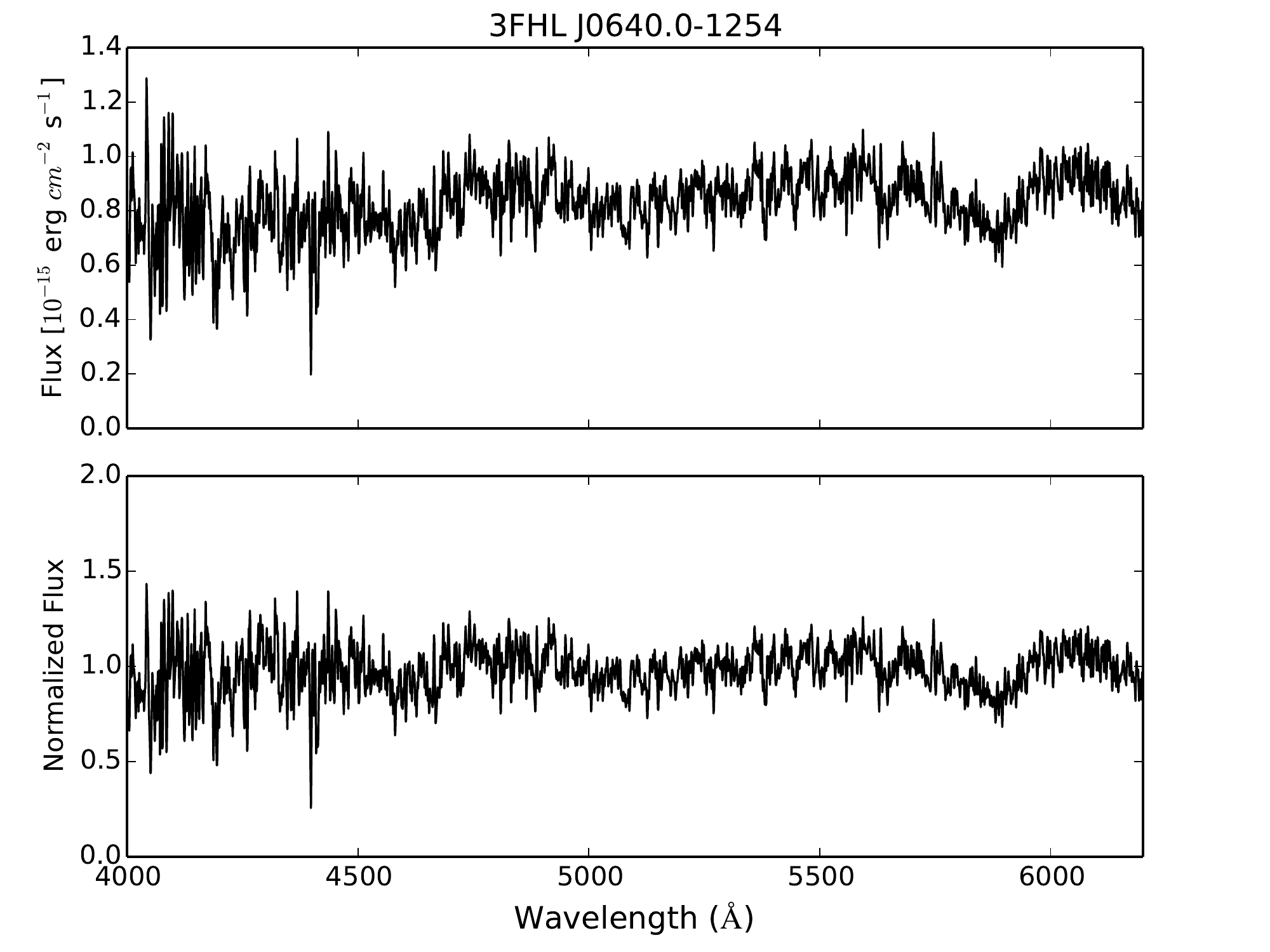}
  \end{minipage}
  \begin{minipage}[b]{.5\textwidth}
  \centering
  \includegraphics[width=0.9\textwidth]{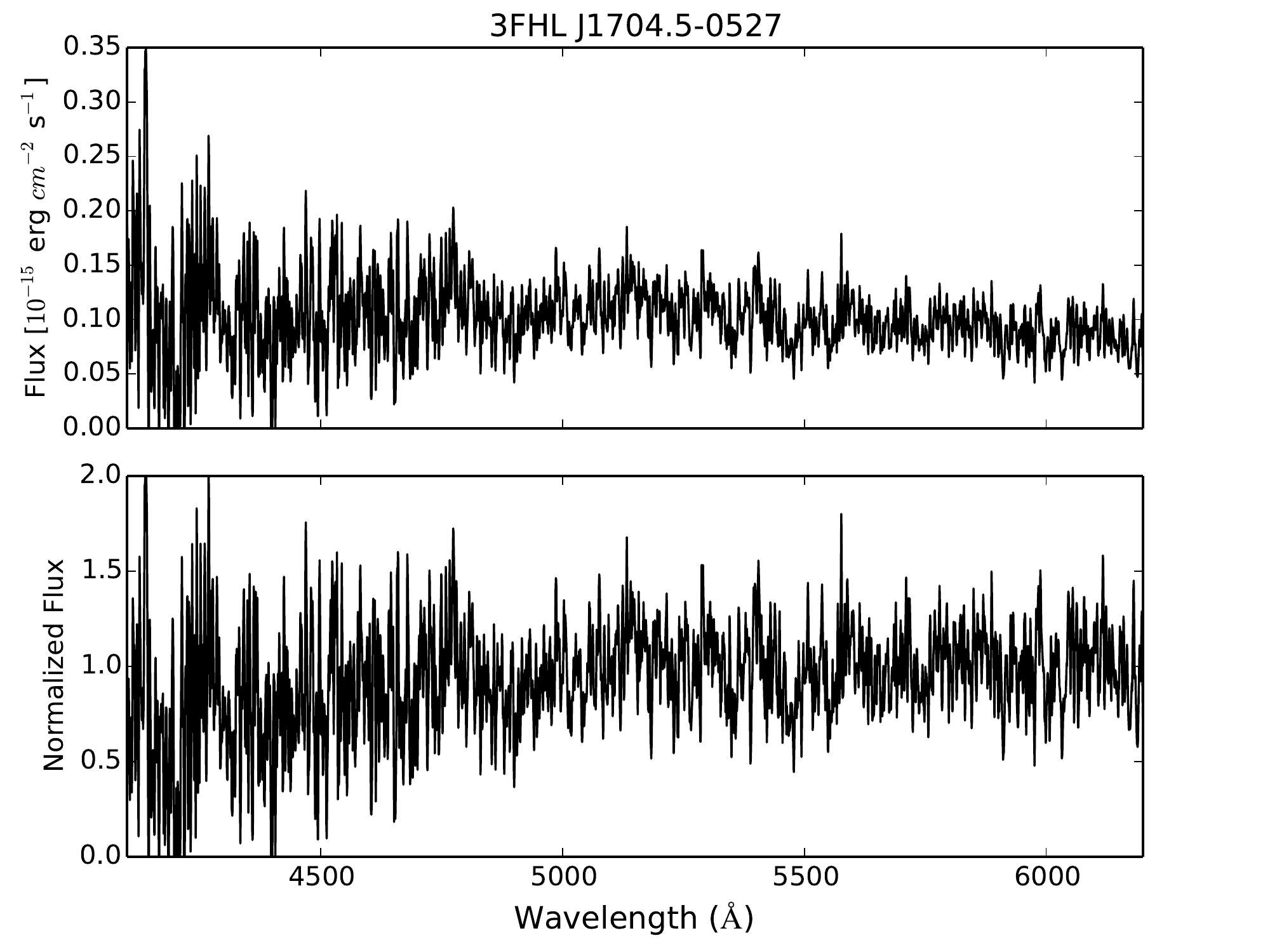}
  \end{minipage}
\caption{Optical spectra of six out of 28 objects in our sample. The spectra have been smoothed for visualization purposes.}\label{fig:spectra2}
\end{figure*}

\begin{figure*}
\begin{minipage}[b]{.5\textwidth}
  \centering
  \includegraphics[width=0.9\textwidth]{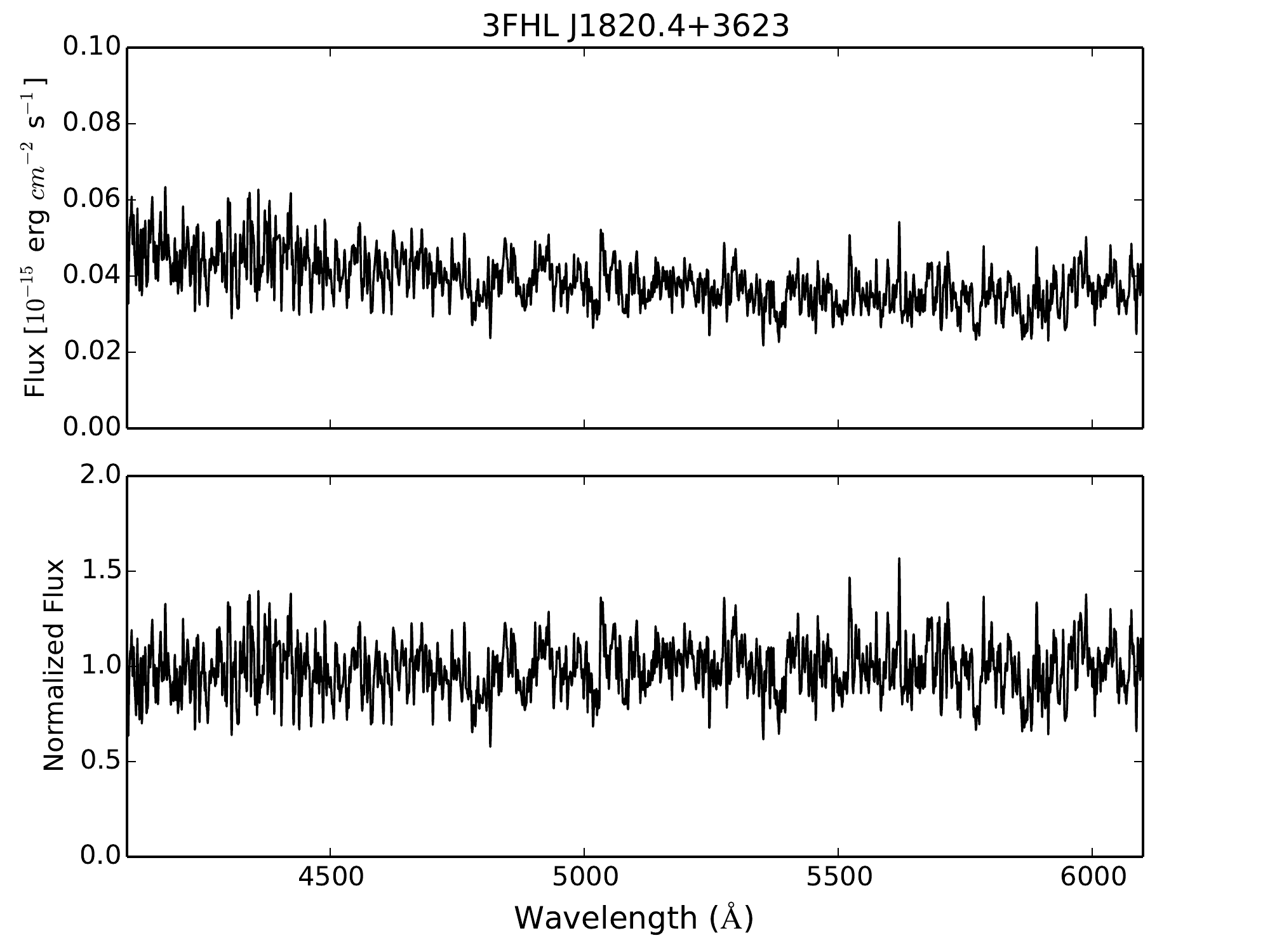}
  \end{minipage}
    \vspace{0.5cm}
\begin{minipage}[b]{.5\textwidth}
  \centering
  \includegraphics[width=0.9\textwidth]{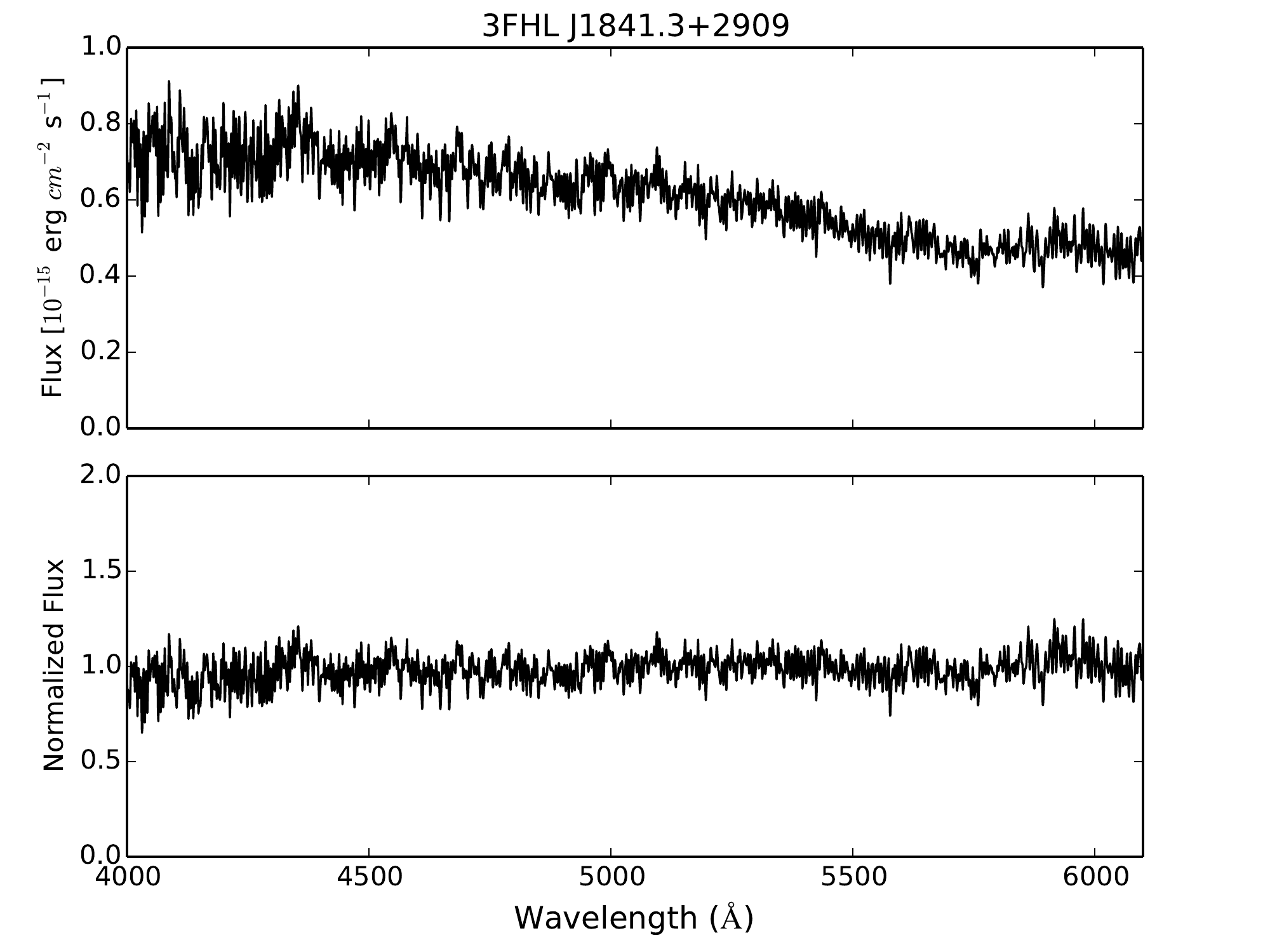}
  \end{minipage}
\begin{minipage}[b]{.5\textwidth}
  \centering
  \includegraphics[width=0.9\textwidth]{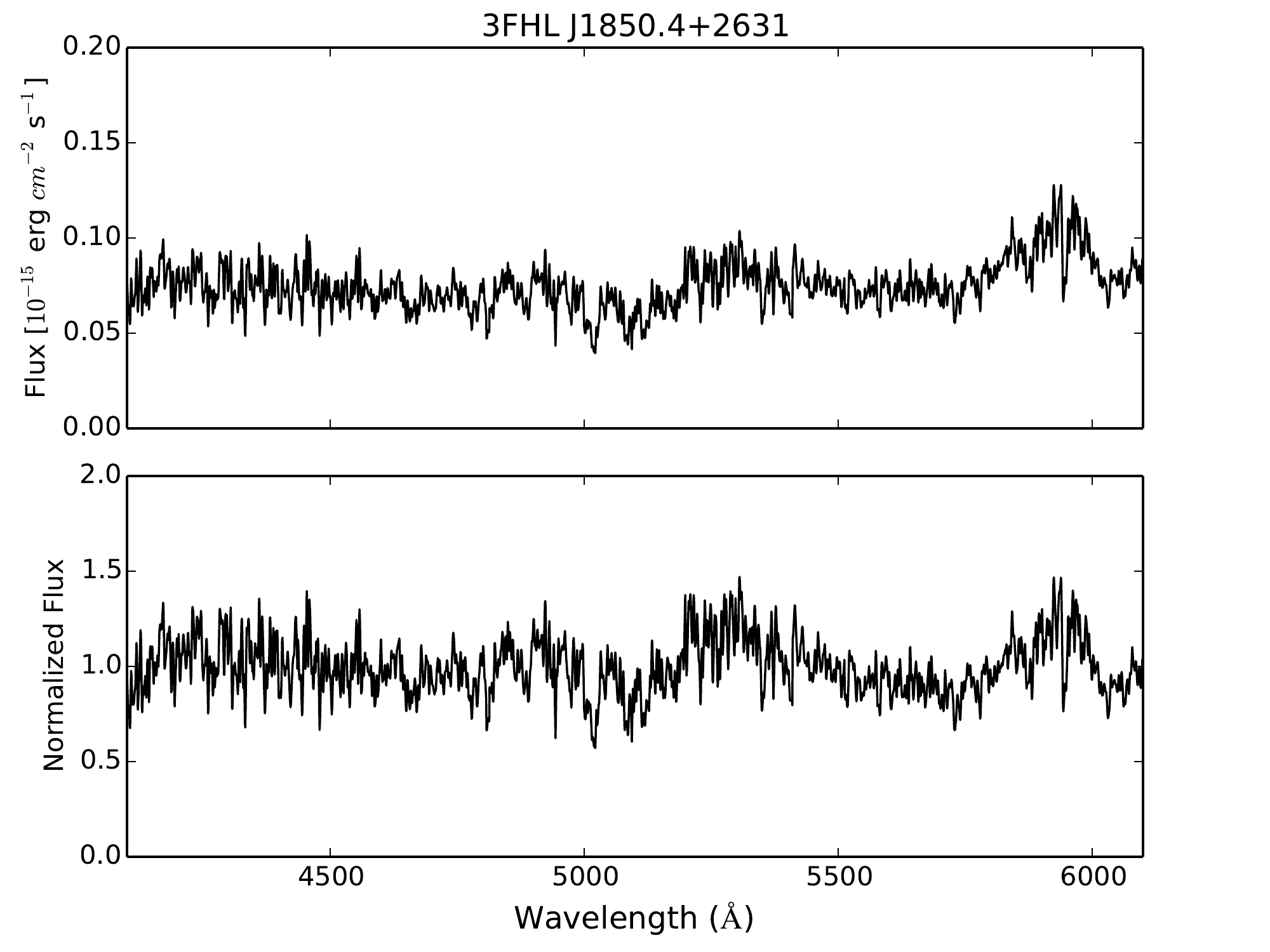}
  \end{minipage}
    \vspace{0.2cm}
\begin{minipage}[b]{.5\textwidth}
  \centering
  \includegraphics[width=0.9\textwidth]{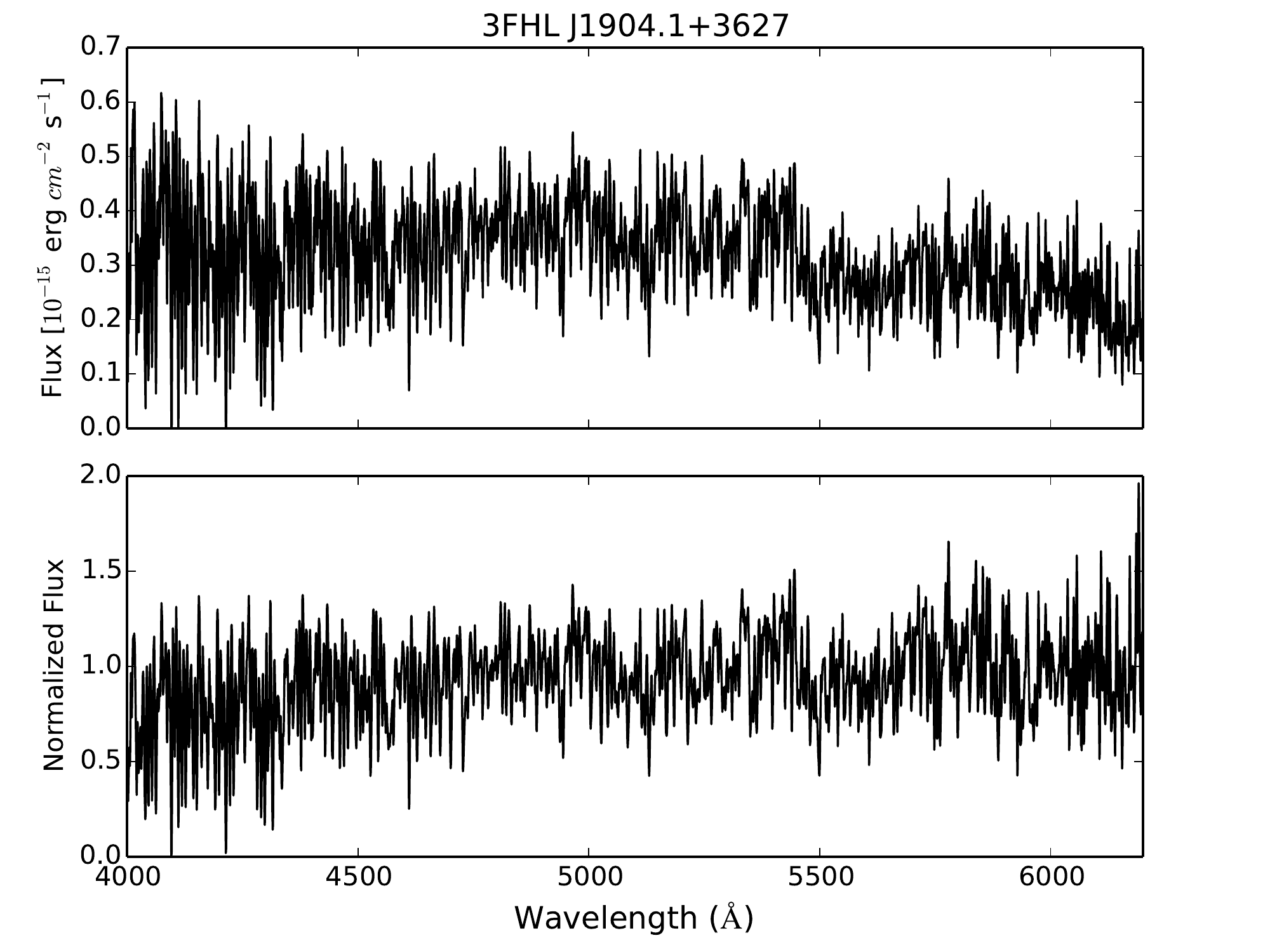}
  \end{minipage}
  \begin{minipage}[b]{.5\textwidth}
  \centering
  \includegraphics[width=0.9\textwidth]{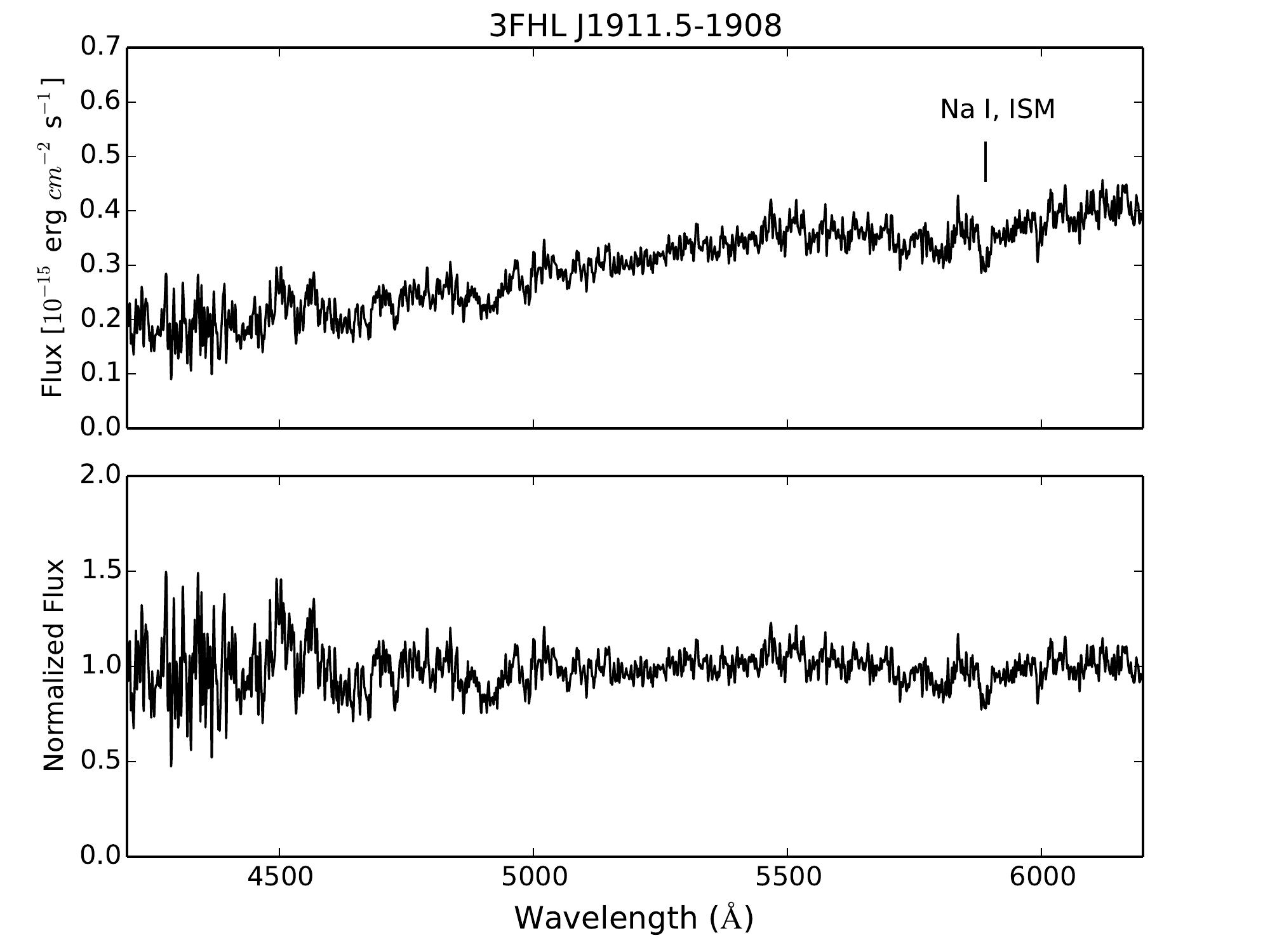}
  \end{minipage}
  \begin{minipage}[b]{.5\textwidth}
  \centering
  \includegraphics[width=0.9\textwidth]{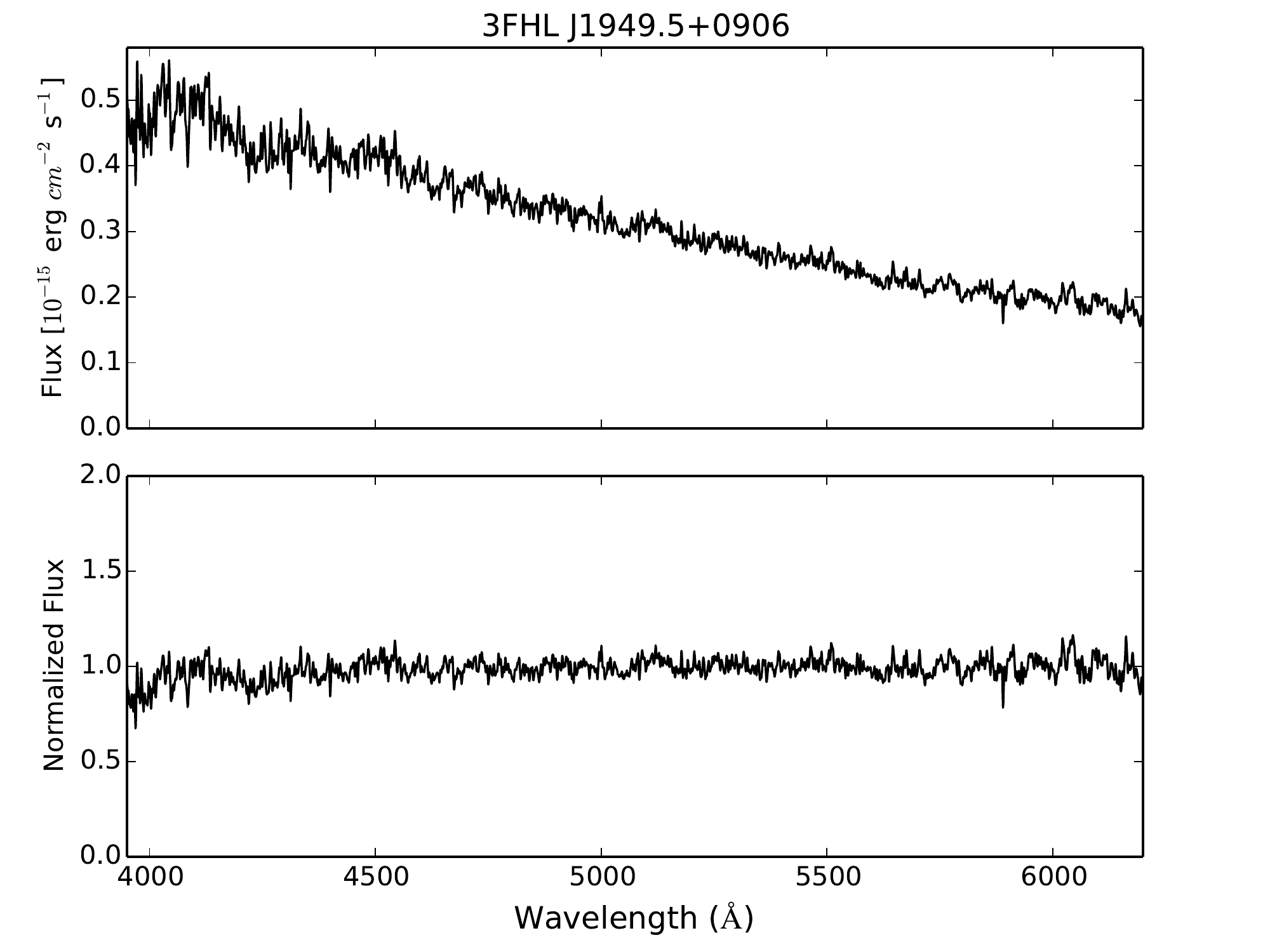}
  \end{minipage}
\caption{Optical spectra of six out of 28 objects in our sample. The spectra have been smoothed for visualization purposes.}\label{fig:spectra3}
\end{figure*}

\begin{figure*}
\begin{minipage}[b]{.5\textwidth}
  \centering
  \includegraphics[width=0.9\textwidth]{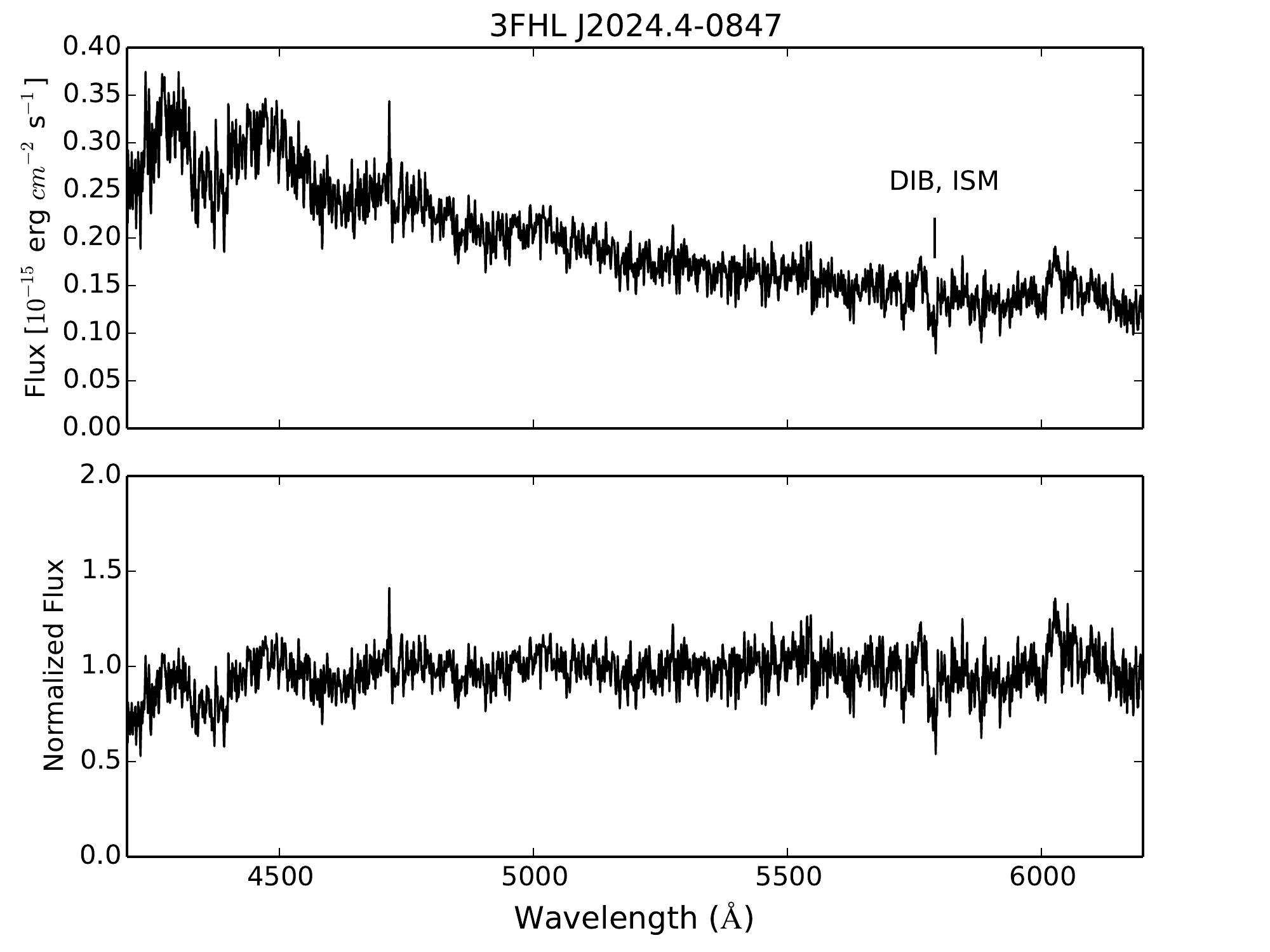}
  \end{minipage}
    \vspace{0.5cm}
\begin{minipage}[b]{.5\textwidth}
  \centering
  \includegraphics[width=0.9\textwidth]{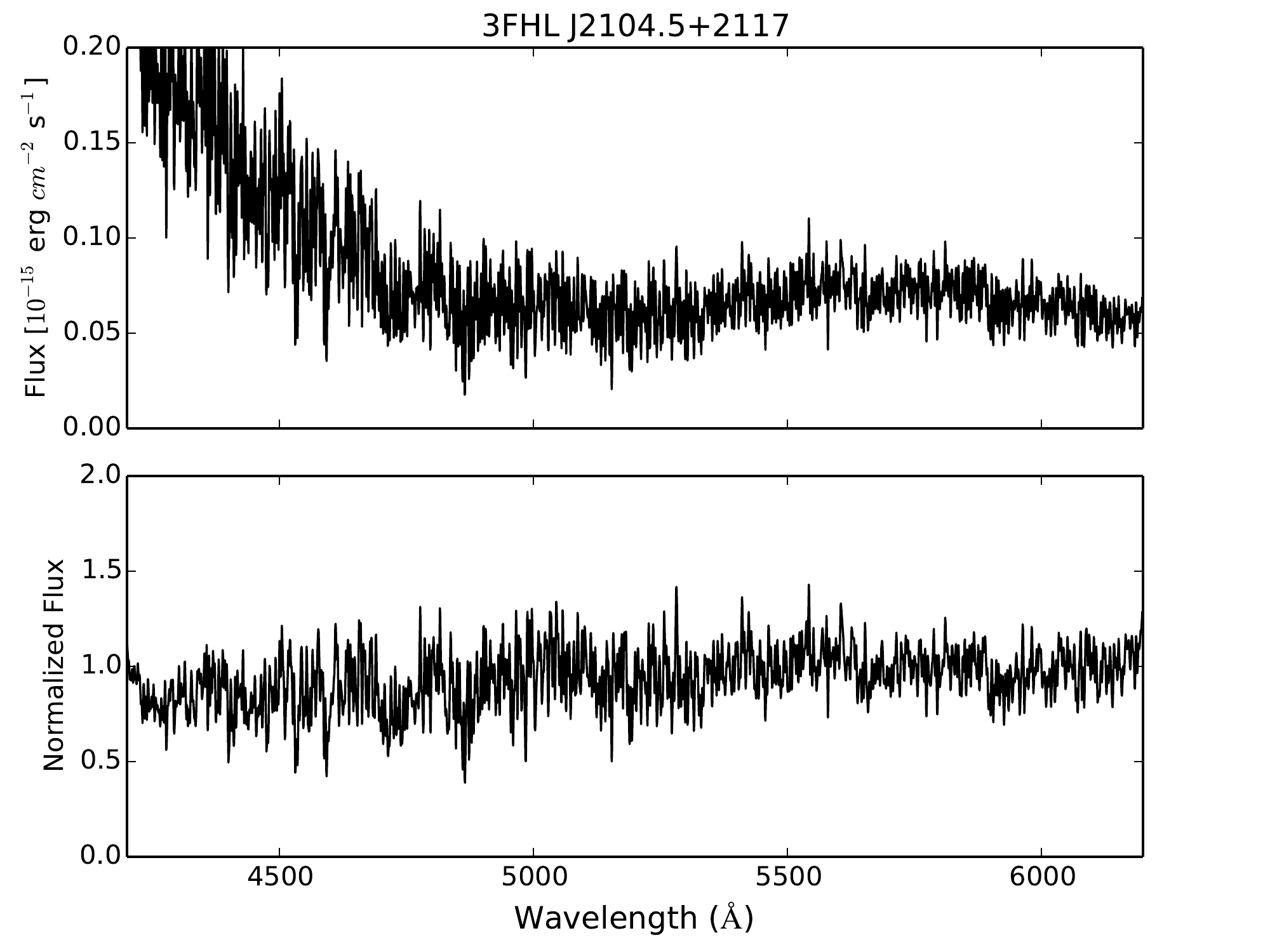}
  \end{minipage}
    \vspace{0.2cm}
\begin{minipage}[b]{.5\textwidth}
  \centering
  \includegraphics[width=0.9\textwidth]{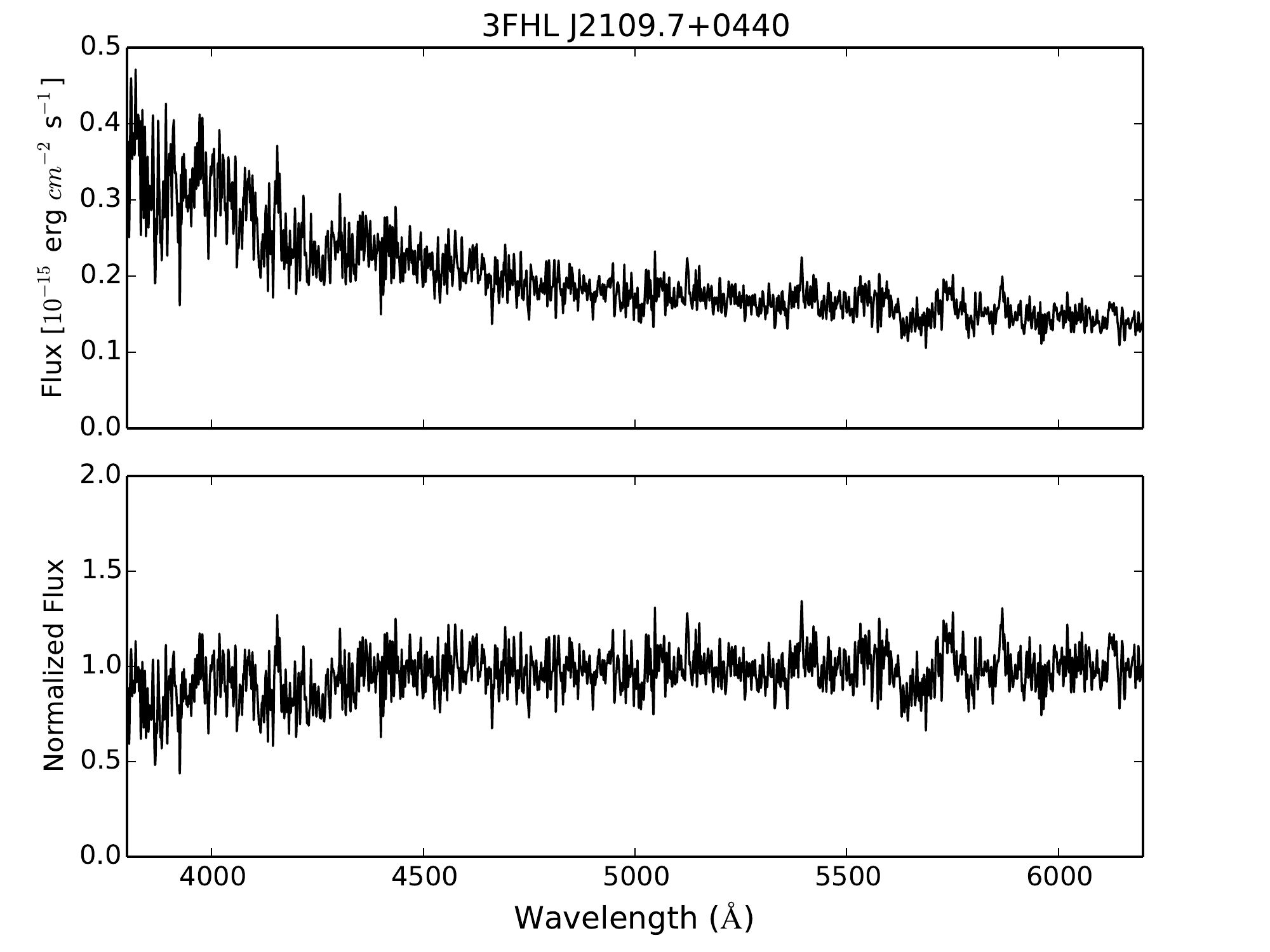}
  \end{minipage}
  \begin{minipage}[b]{.5\textwidth}
  \centering
  \includegraphics[width=0.9\textwidth]{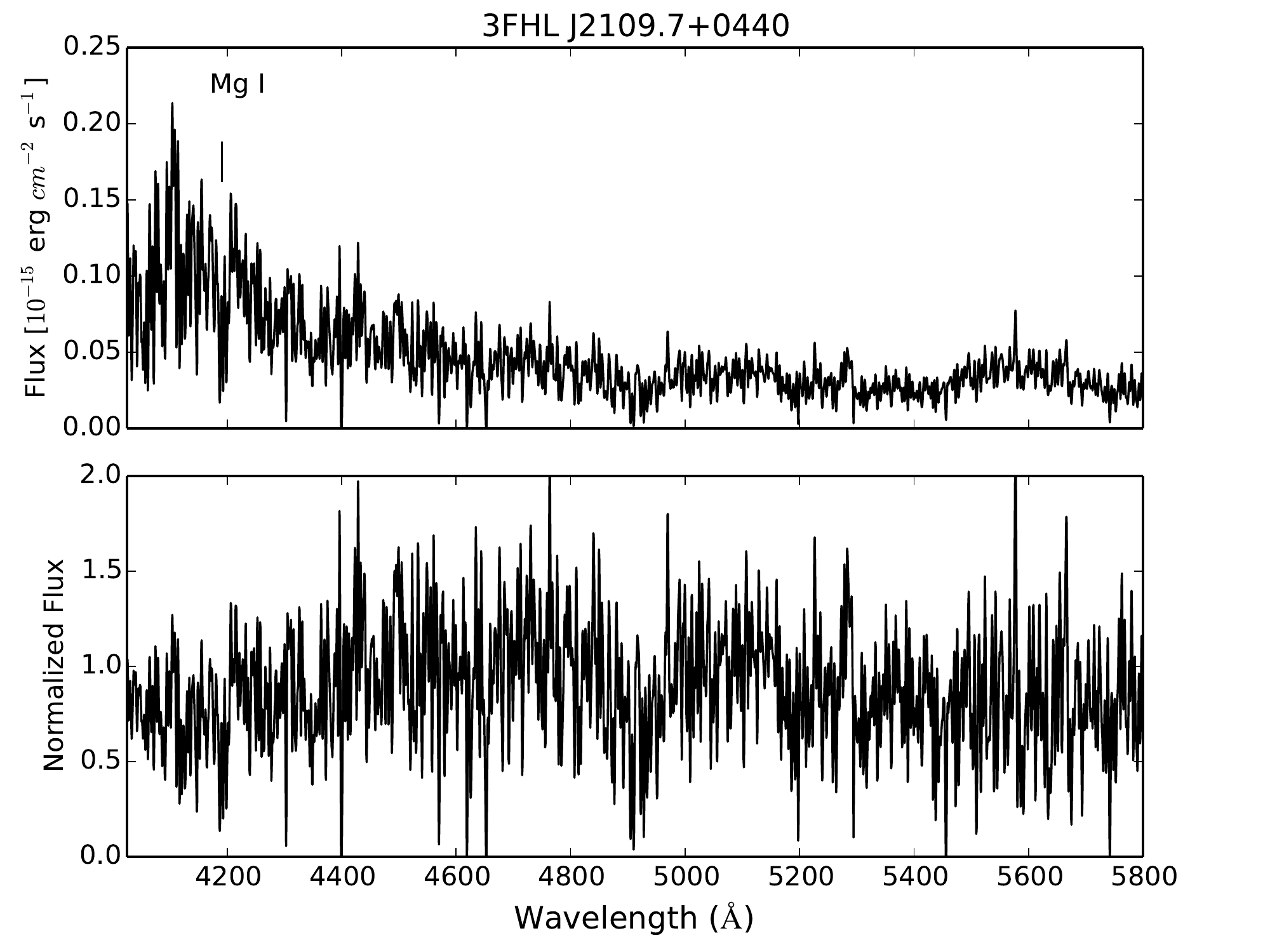}
  \end{minipage}
  \begin{minipage}[b]{.5\textwidth}
  \centering
  \includegraphics[width=0.9\textwidth]{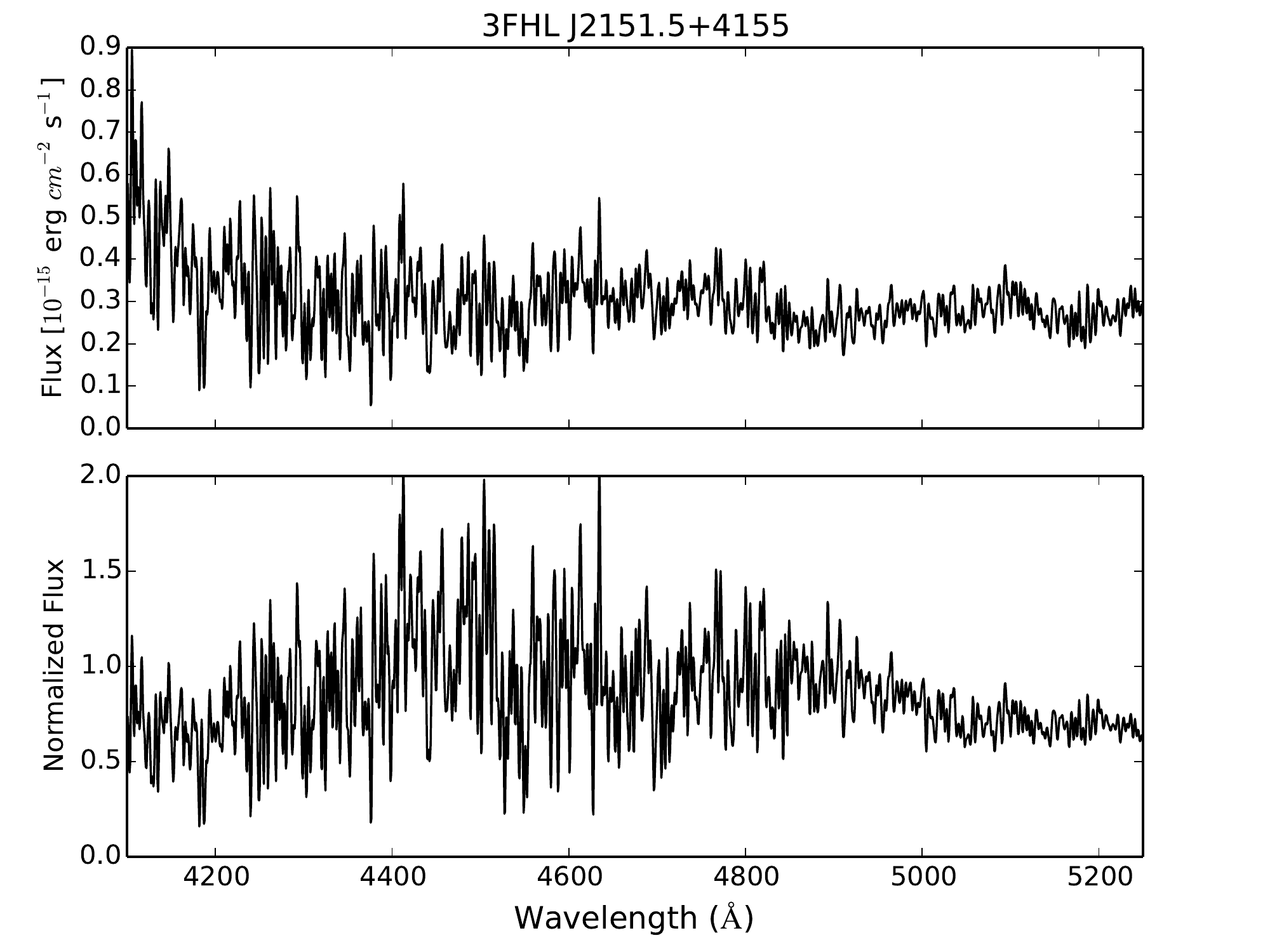}
  \end{minipage}
  \begin{minipage}[b]{.5\textwidth}
  \centering
  \includegraphics[width=0.9\textwidth]{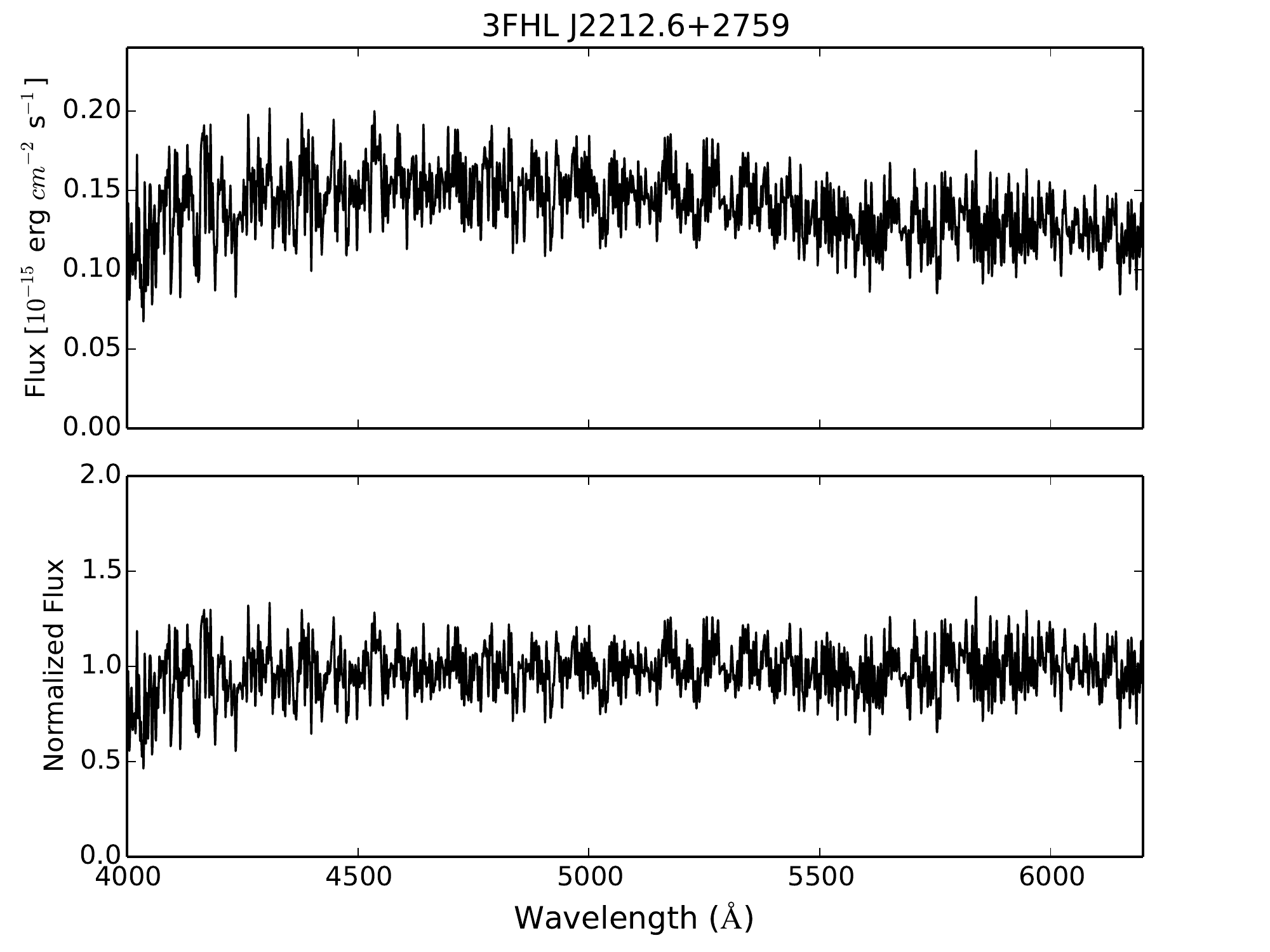}
  \end{minipage}
\caption{Optical spectra of six out of 28 objects in our sample. The spectra have been smoothed for visualization purposes.}\label{fig:spectra4}
\end{figure*}

\begin{figure*}
    \vspace{0.5cm}
\begin{minipage}[b]{.5\textwidth}
  \centering
  \includegraphics[width=0.9\textwidth]{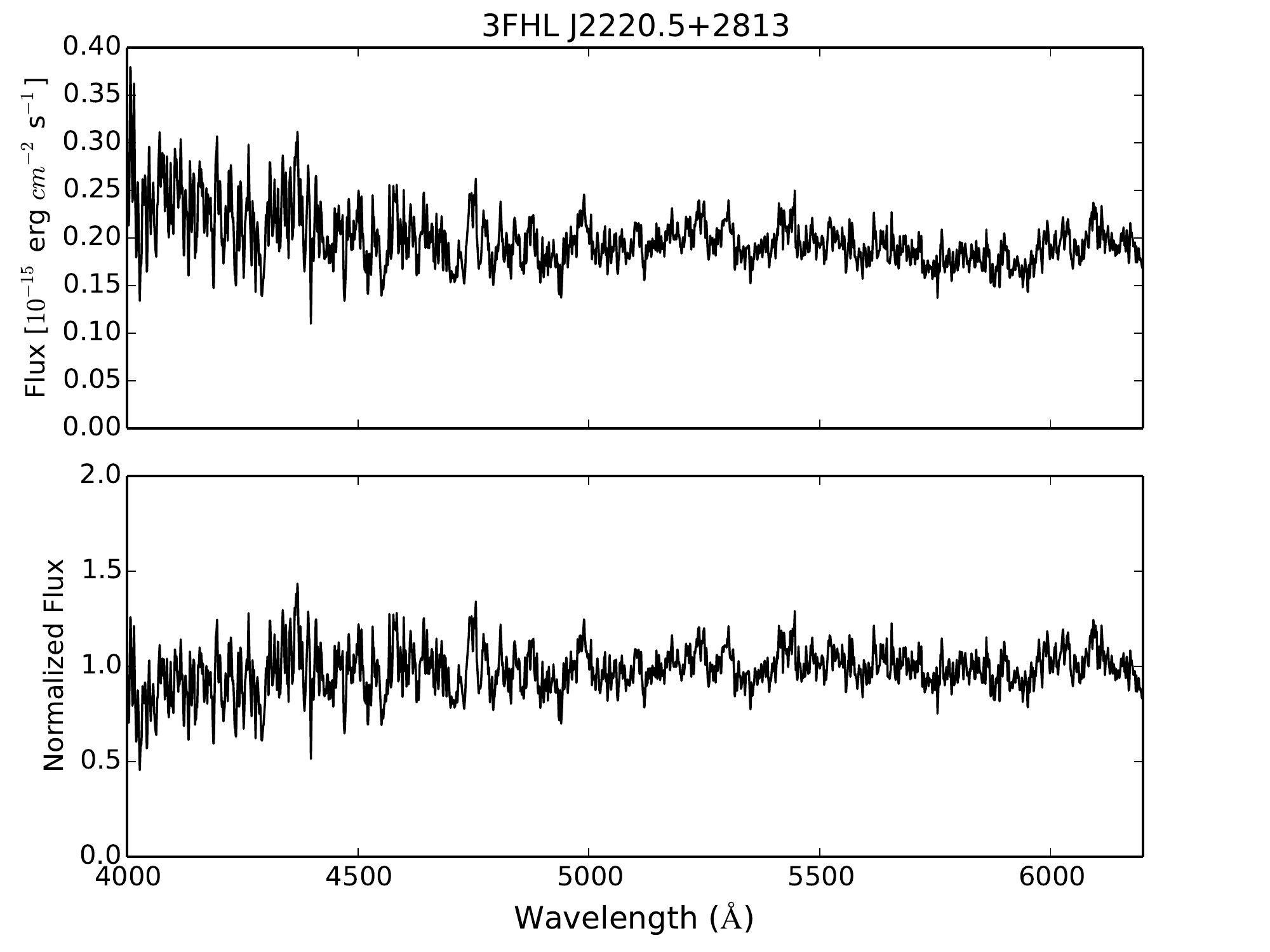}
  \end{minipage}
\begin{minipage}[b]{.5\textwidth}
  \centering
  \includegraphics[width=0.9\textwidth]{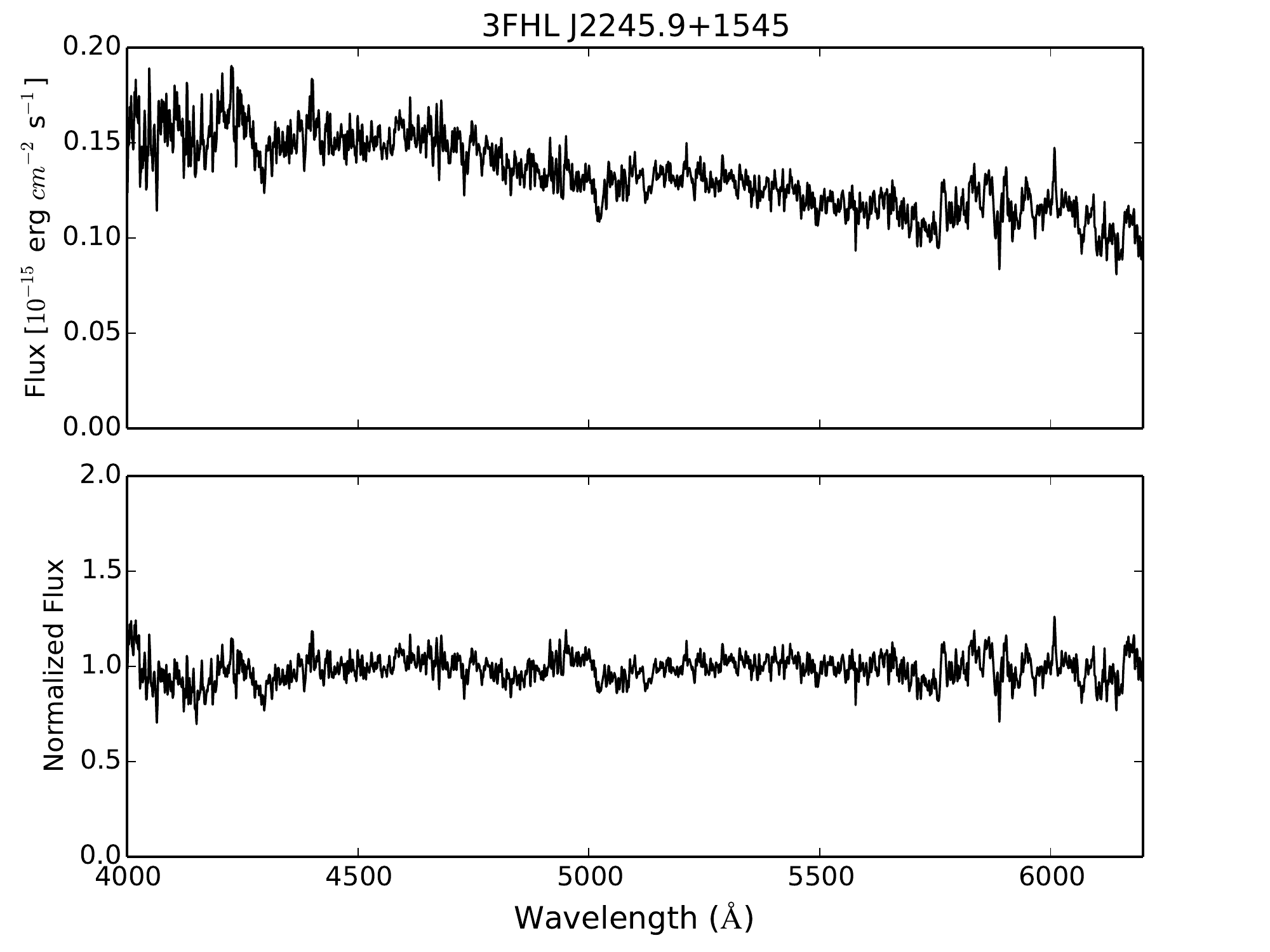}
  \end{minipage}
    \vspace{0.2cm}
\begin{minipage}[b]{.5\textwidth}
  \centering
  \includegraphics[width=0.9\textwidth]{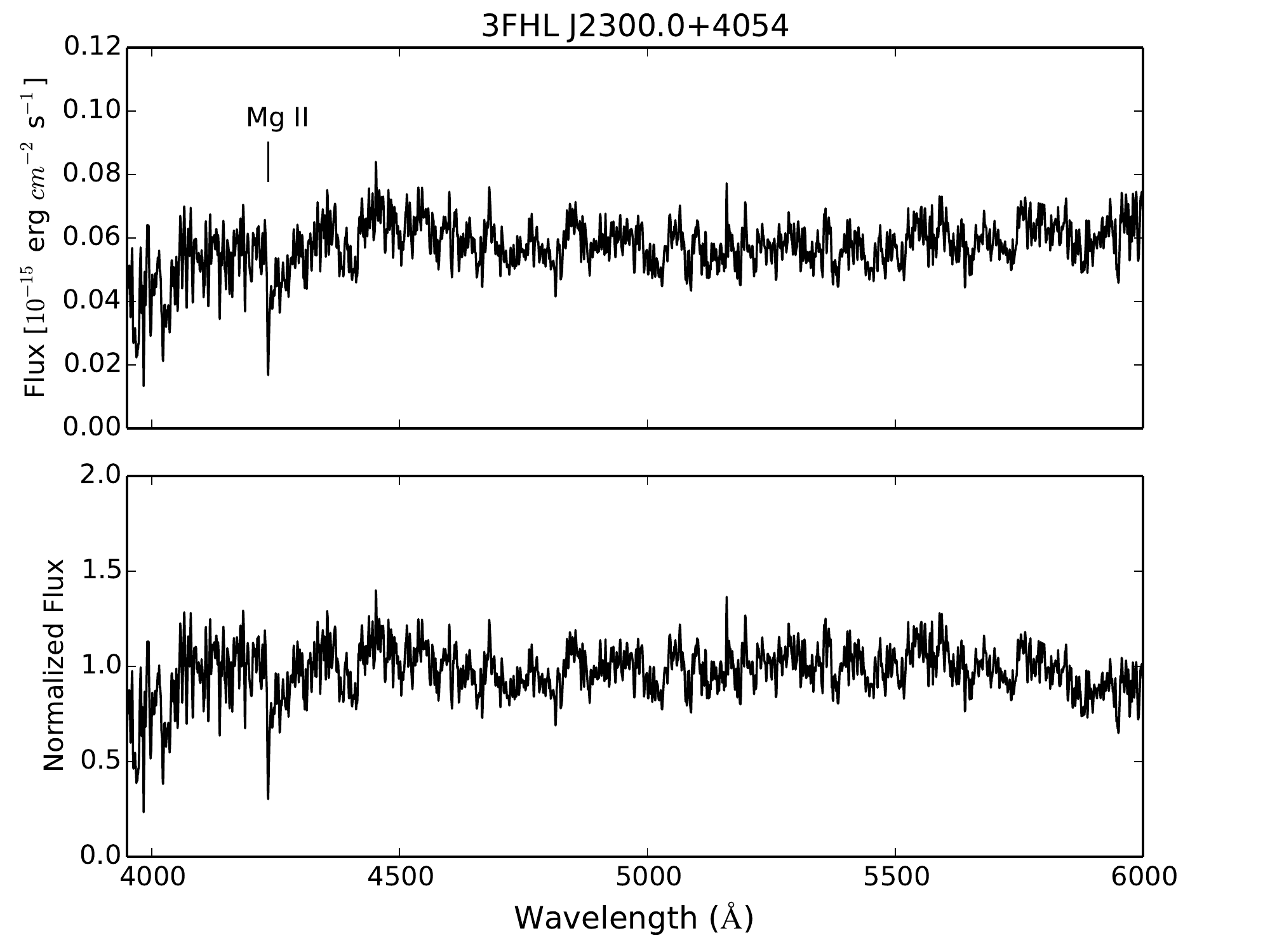}
  \end{minipage}
  \begin{minipage}[b]{.5\textwidth}
  \centering
  \includegraphics[width=0.9\textwidth]{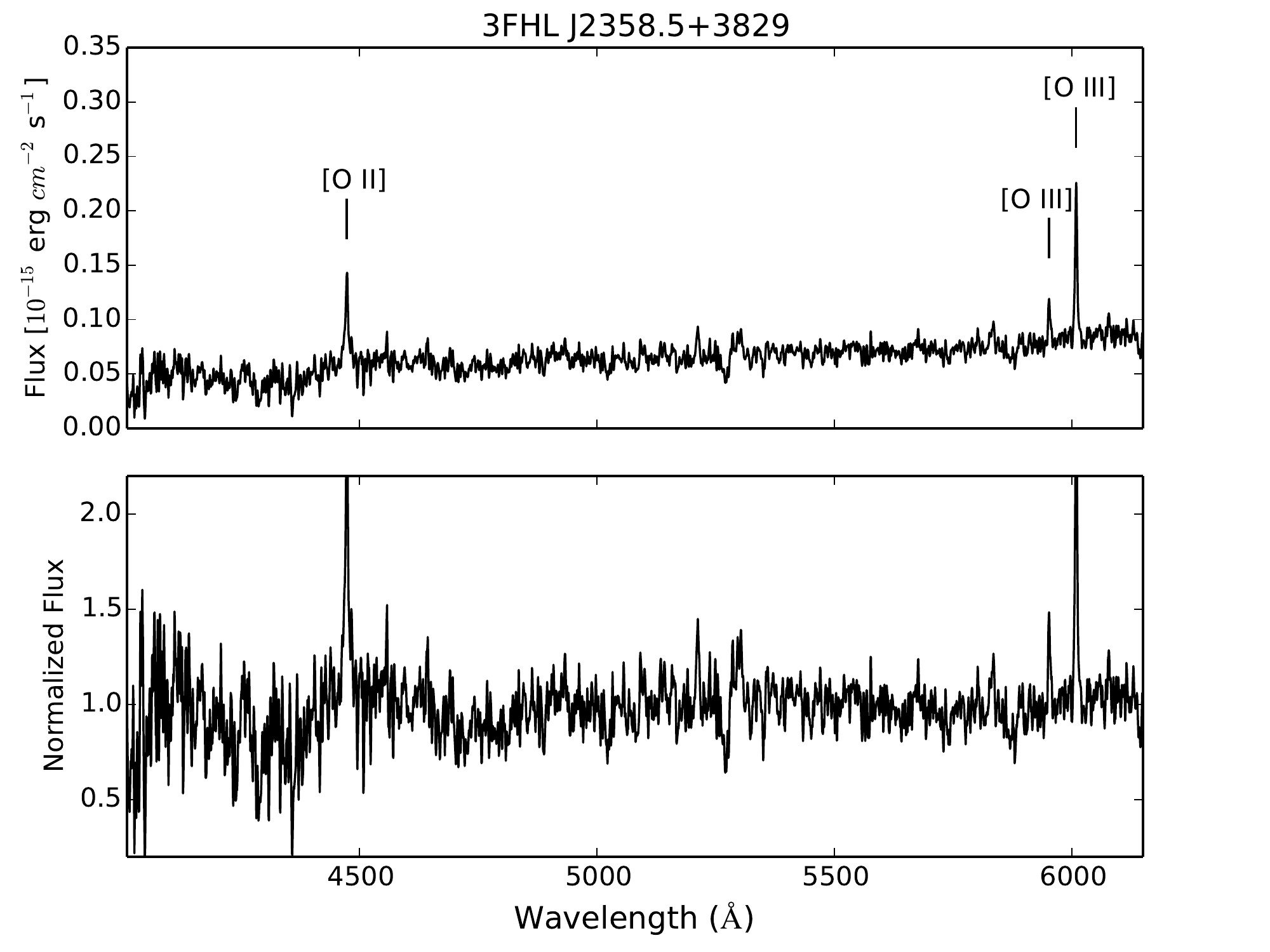}
  \end{minipage}
\caption{Optical spectra of four out of 28 objects in our sample. The spectra have been smoothed for visualization purposes.}\label{fig:spectra_last}
\end{figure*}

The flux-calibrated spectra, together with the normalized spectra, are shown in Figures \ref{fig:spectra1}--\ref{fig:spectra_last}\footnote{The spectra are also available online, both in FITS and eps format, at \url{https://clemson.box.com/s/uu1hk6g4qy0ow9j4nst4nifm3mmrs19f}}. The normalized spectra are obtained dividing the flux-calibrated spectrum by a power law fit of the continuum.

All spectra have been visually inspected to identify candidate emission and absorption features. In order to deem a feature reliable, we verified its existence in each of the individual spectra which are combined in the final one.

As a first result of this follow-up program, we are able to classify all the sources in our sample: 27 out of 28 objects are BL Lacs, while the remaining one is a FSRQ. We determine a redshift for three of these objects and a lower limit on $z$ for other four sources. 

Finally, 21 objects are classified as featureless BL Lacs. These are sources where a 4m instrument does not allow to reach a S/N high enough to observe faint features, but represent a potential candidate for further observations with 8m or 10m telescopes. Notably, while we find no clear feature in 75\% of the objects in our sample, in a recent work by \citep{paiano17} only 4 out of 20 blazars (20\% of their sample) observed with the 10m Gran Telescopio Canarias (GTC) are found to be featureless.

\subsection{Comments on individual sources}\label{sec:individual}
\begin{itemize}
\item 3FHL J0031.2+0727: this object is a BCU associated with the optical source SDSS J003119.71+072453.5 and to the radio source NVSS J003119+072456.  We observe six narrow absorption features, at 4374\,\AA\ (EW=1.0\,\AA), 4749\,\AA\ (EW=1.0\,\AA), 4773\,\AA\ (EW=1.0\,\AA), 5133\,\AA\ (EW=2.4\,\AA), 5146\,\AA\ (EW=2.0\,\AA) and 5236\,\AA\ (EW=1.9\,\AA). The first three features are associated with Fe I lines at 2382\,\AA, 2586\,\AA\ and 2599\,\AA, respectively, while the latter three are associated with the Mg I lines at 2797\,\AA\ and 2803\,\AA\ and to the Mg II line at 2852\,\AA, respectively. Since these absorption features can in principle be associated with material on the line of sight between us and SDSS J003119.71+072453.5, our redshift measurement should be treated as a lower limit, $z$$>$0.836. The source is classified as a BL Lac.
\item 3FHL J0040.3+4049: associated with the radio source NVSS J004013+405005, this source was classified as a BCU. We do not detect any significant emission or absorption feature in the source spectrum: the object is therefore classified as a BL lac, with no redshift information. 
\item 3FHL J0127.2+0325: BCU associated with the radio source NVSS J012713+032259. The spectrum is consistent with a featureless power law and the object is classified as a BL Lac, with no constraint on its redshift. This object has been recently observed also by \citet{pena17} using the 4.1m Southern Astrophysical Research Telescope (SOAR) at Cerro Pach\'{o}n, Chile. They confirm the lack of features in the source spectrum and the BL Lac classification.
\item 3FHL J0134.4+2638: this source is a BCU associated with the optical source SDSS J013428.19+263843.0 and with the radio source NVSS J013427+263842. 
We observe a prominent emission lines at 4400\,\AA\ (EW=64\,\AA): the source is therefore classified as a FSRQ. 
The observed feature can be associated with Mg II at 2800\,\AA, which implies that the source redshift is $z$=0.571.
\item 3FHL J0241.3+6543: BCU associated with the radio source NVSS J024121+654311. Based on the optical spectrum, the source is BL Lac, with a power law dominated spectrum. We observe an absorption feature at $\sim$4600\,\AA: more precisely, the feature is a doublet, with a peak at 4602\,\AA\ and the other at 4609\,\AA. These wavelengths are consistent with being those of the Mg II doublet (rest-frame wavelength  $\lambda_{\rm r}$=2797--2803\,\AA), therefore suggesting a redshift lower limit $z$$>$0.645 for 3FHL J0241.3+6543. 
\item 3FHL J0305.2--1609: BCU associated with the radio source PKS 0302--16. The source is a BL Lac, and we find evidence of a narrow emission feature at 4886\,\AA\ (EW=1\,\AA), that can be linked to [OII] at $\lambda_{\rm r}$=3727\,\AA, thus implying a redshift $z$=0.311. This emission feature, as well as several absorption features that we do not observe due to low S/N, has been detected also by \citet{paiano17} using the GTC 10m telescope: their redshift measurement is in perfect agreement with ours, $z$=0.311.
\item 3FHL J0353.4+8256: BCU associated with the radio source NVSS J035309+825631. The source spectrum is well fitted by a power law and we classify it as featureless BL Lac.
\item 3FHL J0420.2+4011: BCU associated with the radio source NVSS J042013+401122. This object is a featureless BL Lac. 
\item 3FHL J0431.8+7403: BCU associated with the radio source NVSS J043145+740327. This object is a featureless BL Lac. 
\item 3FHL J0434.7+0921: BCU associated with the radio source NVSS J043440+092348. Featureless BL Lac. 
\item 3FHL J0640.0--1254:  BCU associated with the radio source NVSS J064007--125315. This object has been classified as a potential BL Lac candidate on the basis of its X-ray and infrared WISE properties by \citet{massaro13}. The emission is dominated by a non-thermal power law: no clear emission or absorption lines are detected, and the object is thus classified as a BL Lac with no $z$ information. 
\item 3FHL J1704.5--0527: this object is classified as unassociated in the 3FHL catalog. In the \xrt\ image we find a single bright X-ray counterpart within the 3FHL error box (Kaur et al. in prep. 2018; see also Figure \ref{fig:xrt_unassociated}, top left panel). The X-ray counterpart is associated with the radio source NVSS J170433--052839. We classify the object as a featureless BL Lac. The same result was recently obtained by \citep{paiano17}, which also set a lower limit $z$$>$0.7 on this object, based on the minimum equivalent width measured in the spectrum \citep[see][for an extended description of this technique]{paiano17b}.
\item 3FHL J1820.4+3623: BCU associated with the radio source NVSS J182021+362343. Based on the optical spectrum, this is a featureless BL LAC. 
\item 3FHL J1841.3+2909: BCU associated with the radio source NVSS J184121+290945. This object has been classified as a potential BL Lac candidate on the basis of its X-ray and infrared WISE properties by \citet{massaro13}. The optical spectrum is a pure power law and the source is therefore classified as a featureless BL Lac.
\item 3FHL J1850.4+2631: BCU associated with the radio source NVSS J185023+263151. The optical spectrum is consistent with a featureless BL Lac one. 
\item 3FHL J1904.1+3627: BCU associated with the radio source NVSS J190411+362700. Based on the optical spectrum, this is a featureless BL LAC. 
\item 3FHL J1911.5--1908: BCU associated with the radio source NVSS J191129--190823. The source is classified as a featureless BL Lac. We find an absorption feature at $\sim$5890\,\AA, which is associated with Na I in the Galactic interstellar medium (ISM). 
\item 3FHL J1949.5+0906: BCU associated with the infrared source 2MASSJ19493419+0906537. The source is classified as a BL Lac, and does not show any clear evidence of features. 
\item 3FHL J2024.4--0847: BCU associated with the radio source NVSS J202429--084804. The spectrum is dominated by non thermal emission and the source is a featureless BL Lac. We find an absorption feature at $\sim$5790\,\AA, which is associated with a diffuse interstellar band  in the interstellar medium (ISM). 
This source has been recently observed by \citet{pena17} with the SOAR 4.1\,m telescope: they were also able to only identify the ISM absorption feature.
\item 3FHL J2104.5+2117: another unassociated source in the 3FHL catalog, for which we find a bright X-ray counterpart through a \xrt\ observation (Figure \ref{fig:xrt_unassociated}, top right panel). The X-ray counterpart is associated with the radio source NVSS J210415+211805. The source is a BL Lac and shows no significant emission or absorption feature. 
\item 3FHL J2109.7+0440: unassociated source in the 3FHL catalog, for which we find a counterpart through a \xrt\ observation (Figure \ref{fig:xrt_unassociated}, central left panel). Within the error box of the 3FHL source there is a single bright X-ray counterpart, which is associated with the radio source NVSS J210939+044000. The optical spectrum is consistent with a BL Lac one, with no clear feature observed either in emission or in absorption. 
\item 3FHL J2115.2+1218: unassociated source in the 3FHL catalog, for which we find a counterpart through a \xrt\ observation (Figure \ref{fig:xrt_unassociated}, central right panel), which is associated with the radio source NVSS J211522+121802. The spectrum is a BL Lac one and we identify a candidate absorption feature at $\lambda_{\rm obs}$$\sim$4190\,\AA. Associating this feature to intervening Mg II at $\lambda_{\rm r}$=2800\,\AA\ we can get a redshift lower limit, $z$$>$0.496. This same source has been identified also by \citet{paiano17} with the 10m GTC telescope: they claim to find several absorption features, the most prominent one is the same feature at 4191\,\AA\ ($z$$>$0.497) which we find in our spectrum.
\item 3FHL J2151.5+4155: the last unassociated source from the 3FHL catalog in our sample, this object is associated, through the \xrt\ image (Figure \ref{fig:xrt_unassociated}, bottom left panel), with the radio source NVSS J215122+415632. Based on its optical spectrum, this source is a featureless BL Lac.
\item 3FHL J2212.6+2759:  BCU associated with the radio source NVSS J221239+275937. Featureless BL Lac.
\item 3FHL J2220.5+2813: BCU associated with the radio source NVSS J222028+281355. The source optical spectrum is dominated by the non-thermal emission and the object is a featureless BL Lac.
\item 3FHL J2245.9+1545: BCU associated with the radio source NVSS J224604+154437. No clear emission or absorption lines are observed in the optical spectrum of this object, which we classify as a BL Lac. 
This source has been analyzed also by \citet{paiano17}: they also find this source to be featureless, and set a lower limit on the redshift, $z$$>$0.7.
\item 3FHL J2300.0+4054: BCU associated with the radio source NVSS J230012+405224. The optical spectrum of the source is a consistent with a BL Lac one. 
We detect a narrow absorption feature at $\lambda_{\rm obs}$=4235\,\AA\ (EW=2.8\,\AA). If this absorption line is caused by Mg II, we can set a redshift lower limit $z$$>$0.513. 
\item 3FHL J2358.5+3829: BCU associated with the radio source NVSS J235825+382857. We detect several emission lines, at 4473\,\AA\ (EW=6.4\,\AA), 5992\,\AA\ (EW=3.2\,\AA),  and 6009\,\AA\ (EW=8.6\,\AA). The first line is associated with [OII] at 3727\,\AA, while the other two emission features correspond to the [OIII] emission lines at 4959\,\AA\ and 5007\,\AA, respectively: the redshift of the source is therefore $z$=0.200, and the object is classified as a  BL Lac. 
\end{itemize}


\begingroup
\renewcommand*{\arraystretch}{1.8}
\begin{table*}
\centering
\scalebox{1}{
\begin{tabular}{cccc}
\hline
\hline
3FHL Name & S/N & Type & $z$\\
\hline
J0031.2+0727 & 25 & BL Lac & $>$0.836\\
J0040.3+4049 & 19 & BL Lac & --\\
J0127.2+0325 & 12 & BL Lac & --\\
J0134.4+2638 & 27 & FSRQ & 0.571$\pm$0.002\\
J0241.3+6543 & 19 & BL Lac & $>$0.645\\
J0305.2$-$1609 & 22 & BL Lac & 0.311$\pm$0.001\\
J0353.4+8256 & 20 & BL Lac & --\\
J0420.2+4011 & 25 & BL Lac & --\\
J0431.8+7403 & 52 & BL Lac & --\\
J0434.7+0921 & 51 & BL Lac & --\\
J0640.0$-$1254 & 20 & BL Lac & --\\
J1704.5$-$0527 & 10 & BL Lac & --\\
J1820.4+3623 & 15 & BL Lac & --\\
J1841.3+2909 & 33 & BL Lac & --\\
J1850.4+2631 & 25 & BL Lac & --\\
J1904.1+3627 & 10 & BL Lac & --\\
J1911.5$-$1908 & 32 & BL Lac & --\\
J1949.5+0906 & 36 & BL Lac & --\\
J2024.4$-$0847  & 26 & BL Lac & --\\
J2104.5+2117 & 16 & BL Lac & --\\
J2109.7+0440 & 17 & BL Lac & \\
J2115.2+1218 & 21 & BL Lac & $>$0.496\\
J2151.5+4155 & 26 & BL Lac & --\\
J2212.6+2759 & 19 & BL Lac & --\\
J2220.5+2813 & 22 & BL Lac & --\\
J2245.9+1545 & 26 & BL Lac & --\\
J2300.0+4054 & 24 & BL Lac & $>$0.531\\
J2358.5+3829 & 20 & BL Lac & 0.200$\pm$0.001\\
\hline
\hline
\end{tabular}}\caption{Summary of the spectral analysis results. S/N is the spectrum signal-to-noise ratio, Type is the source new classification, $z$ is the redshift value or lower limit, when measurable.}
\label{tab:results}
\end{table*}
\endgroup

\section{Conclusions}\label{sec:concl}
In this work, we report the results of the first part of an optical spectroscopy campaign aimed at increasing the spectral completeness of the 3FHL catalog. We observed 28 3FHL sources with the KOSMOS spectrograph mounted on the 4m Mayall telescope at Kitt Peak: 
23 of these objects are BCUs, i.e., blazars of uncertain classification, while the remaining five are sources unassociated in the 3FHL catalog for which we found a optical-radio counterpart through a follow-up with \xrt\ (Kaur et al. submitted).

We were able to reliably classify all the 28 sources in our sample: 27 objects are BL Lacs, and only one source is a FSRQ (3FHL J0134.4+2638). The fact that the vast majority of sources in our sample are BL Lacs is not unexpected: among the 873 blazars already classified in the 3FHL catalog, 731 BL Lacs (83.7\% of the sample of classified blazars) and 142 FSRQs (16.3\%). This is due to the fact that above 10\,GeV, i.e., in the energy range covered in the 3FHL catalog, BL Lacs are significantly brighter than FSRQs.

Overall, we put a constraint on the redshift of 25\% of the objects in our sample. This relatively low fraction is consistent with those obtained in the vast majority of works based on 4m instruments \citep[10--35\%; see, e.g.,][]{landoni15,ricci15,alvarez16a,pena17}. The fraction of sources for which it is possible to obtain a redshift measurement is instead significantly larger in surveys performed using 8m+ telescopes \citep[60--80\%; see, e.g.,][]{sbaruffati05,sbaruffati06a,paiano17}: for this reason, we plan to continue our spectroscopic campaign using the 8m Gemini-N and -S telescopes. We have been granted five nights of observations with these two instruments through \textit{Fermi} (Cycle 11, proposal ID: 111128, PI S. Marchesi). The observations will take place in 2019 and we plan to observe 40--50 sources.



\subsection*{Acknowledgements}
We acknowledge funding under NASA contract 80NSSC17K0503. We thank Amy Robertson, David Summers and Doug Williams for the help provided during the observing nights, and Marco Landoni for the useful comments on the data reduction and analysis process.

This work made use of data supplied by the UK Swift Science Data Centre at the University of Leicester, as well as of the TOPCAT software \citep{taylor05} for the analysis of data tables.


\bibliographystyle{aa}
\bibliography{3FHL_kp_accepted}


\end{document}